\documentclass[fleqn,usenatbib]{mnras}

\usepackage{newtxtext,newtxmath}
\usepackage[T1]{fontenc}

\DeclareRobustCommand{\VAN}[3]{#2}
\let\VANthebibliography\thebibliography
\def\thebibliography{\DeclareRobustCommand{\VAN}[3]{##3}\VANthebibliography}

\usepackage{graphicx}	
\usepackage{amsmath}	
\usepackage[english]{babel} 
\usepackage[usenames]{color} 
\usepackage[normalem]{ulem}

\newcommand{\figpath}{figs}

\newcommand{\kmps}{\rm km~s\ensuremath{^{-1} }\,}

\newcommand{\Msun}{\rm M\ensuremath{_\odot}\,}

\newcommand{\Oo}{\displaystyle}

\begin{document}
\title[Galaxies with counterrotation in IllustrisTNG]{Extreme kinematic misalignment in IllustrisTNG galaxies: the origin, structure and internal dynamics of galaxies with a large-scale counterrotation}

\author[Khoperskov et al.]{Sergey Khoperskov$^{1,2,3}$\thanks{E-mail: sergey.khoperskov@gmail.com}, Igor Zinchenko$^{4,5}$, Branislav Avramov$^{6}$,  Sergey Khrapov$^{7}$, \newauthor Peter Berczik$^{8,6,5}$, Anna Saburova$^{3,9}$, Marina Ishchenko$^5$, Alexander Khoperskov$^7$, \newauthor Claudia Pulsoni$^{2,10}$, Yulia Venichenko$^{11}$,  Dmitry Bizyaev$^{12,9}$, Alexei Moiseev$^{13}$  \\
$^1$ Leibniz Institut f\"{u}r Astrophysik Potsdam (AIP), An der Sternwarte 16, D-14482, Potsdam, Germany \\
$^2$ Max-Planck-Institut f\"{u}r extraterrestrische Physik, Gie{\ss}enbachstrasse 1, 85748 Garching, Germany \\
$^3$ Institute of Astronomy, Russian Academy of Sciences, 48 Pyatnitskya St., Moscow, 119017, Russia \\
$^4$ Faculty of Physics, Ludwig-Maximilians-Universität, Scheinerstr. 1, 81679 Munich, Germany \\
$^5$ Main Astronomical Observatory, National Academy of Sciences of Ukraine, MAO/NASU, 27 Akad. Zabolotnoho St 03680 Kyiv, Ukraine \\
$^6$ Zentrum für Astronomie der Universität Heidelberg, Astronomisches Rechen-Institut, Mönchhofstr 12-14, D-69120 Heidelberg, Germany \\
$^7$ Volgograd State University, 100 Universitetskii prospect, 400062, Volgograd, Russia \\
$^8$ National Astronomical Observatories and Key Laboratory of Computational Astrophysics, Chinese Academy of Sciences, 20A Datun Rd., \\ Chaoyang District, Beijing 100101, China\\ 
$^{9}$ Sternberg Astronomical Institute, Moscow State University, 119234, Moscow, Russia \\
$^{10}$ Excellence Cluster Universe, Boltzmannstrasse 2, 85748 Garching, Germany \\
$^{11}$ Heidelberg University, Grabengasse 1, 69117 Heidelberg, Germany \\
$^{12}$ Apache Point Observatory and New Mexico State University, P.O. Box 59, Sunspot, NM, 88349-0059, USA \\ 
$^{13}$ Special Astrophysical Observatory of the Russian AS, 369167, Nizhnij Arkhyz, Russia}
\maketitle

\begin{abstract}
Modern galaxy formation theory suggests that the misalignment between stellar and gaseous components usually results from an external gas accretion and/or interaction with other galaxies. The extreme case of the kinematic misalignment is demonstrated by so-called galaxies with counterrotation that possess two distinct components rotating in opposite directions with respect to each other. We provide an in-deep analysis of galaxies with counterrotation from IllustrisTNG100 cosmological simulations. We have found $25$ galaxies with substantial stellar counterrotation in the stellar mass range of $2\times10^{9}-3\times10^{10}$~\Msun. In our sample the stellar counterrotation is a result of an external gas infall happened $\approx 2-8$~Gyr ago. The infall leads to the initial removal of pre-existing gas, which is captured and mixed together with the infalling component. The gas mixture ends up in the counterrotating gaseous disc. We show that $\approx 90\%$ of the stellar counterrotation formed in-situ, in the counterrotating gas. During the early phases of the infall, gas can be found in inclined extended and rather thin disc-like structures, and in some galaxies they are similar to (nearly-)~polar disc or ring-like structures. We discuss a possible link between the gas infall, AGN activity and the formation of misaligned components. In particular, we suggest that the AGN activity does not cause the counterrotation, although it is efficiently triggered by the retrograde gas infall, and it correlates well with the misaligned component appearance. We also find evidence of the stellar disc heating visible as an increase of the vertical-to-radial velocity dispersion ratio above unity in both co- and counterrotating components, which implies the importance of the kinematical misalignment in shaping the velocity ellipsoids in disc galaxies.
\end{abstract}


\section{Introduction}
The galaxy formation is a complex phenomenon that involves the dark matter assembly process on large and small scales, such as galactic mergers and gas accretion/outflows. From basic assumptions, stellar components inherit properties of the star-forming gas which globally follows the potential of the dark matter haloes. Thereby, a certain level of the spatial and kinematical alignment between these components in galaxies is expected. However, nowadays it is clear that the stellar and gaseous components in many galaxies show discrepancies in shape and misaligned kinematics~\citep[see, e.g.,][]{2011MNRAS.417..882D, 2012A&A...538A...8S, 2015ApJ...798....7B, 2015A&A...582A..21B, 2016MNRAS.463..913J, 2019arXiv191204522L, 2019MNRAS.483..458B}. The most spectacular case of kinematical misalignment is illustrated by galaxies with counterrotation that directly indicates a complex galaxy assembly history, which is  crucial for our understanding of galactic formation and evolution.

Disc galaxies with counterrotating components is a distinctive class of objects that reveal cospatial~(stars-stars or stars-gas) components rotating in the opposite directions~\citep{1994AJ....108..456R,2014ASPC..486...51C}. Since the first discovery of stars-gas~\citep{1987ApJ...318..531G} and stars-stars counterrotation~\cite{1992ApJ...394L...9R}, the number of galaxies with confirmed counterrotating components is gradually increasing over the years~\citep{1994ApJ...432..575M,1995Natur.375..661C,1996ApJ...458L..67B,2009ApJ...694.1550S,2011MNRAS.412L.113C,2013ApJ...769..105K,2015MNRAS.452....2K,2015A&A...581A..65C,2016MNRAS.461.2068K,2018A&A...616A..22P,2019ApJS..244....6S,2020A&A...634A.102P}. It is also worth mentioning a recent discovery of a counterrotating structure in the Milky Way revealing the latest major merger episode took place $\approx 9-10$~Gyr ago~\citep{2018ApJ...863L..28M, 2018ApJ...863..113H}, which likely contributed to the disc heating~\citep{2019A&A...632A...4D}. Therefore, the counterrotation could be a widespread phenomenon and it is possible that most of galaxies actually contain counterrotating components with the mass and luminosity below the observational thresholds.

Some attempts have been made to explain the appearance of such peculiar kinematics by internal processes, e.g. by resonance capturing~\citep{2000MNRAS.319....1T}, dissolving bars~\citep[see, e.g.,][]{1994ApJ...420L..67E}, or increasing of retrograde stars amount due to the angular momentum exchange with bar~\citep{1991A&A...252...75P}. However,  significant age difference of two counterrotating stellar discs in the spiral galaxy NGC~4138~\citep{2014A&A...570A..79P} and very different chemical compositions of the components in NGC~448~\citep{2019A&A...623A..87N} imply that the counterrotating components are more likely assembled from gas accreted on retrograde orbits. This is supported by the discovery of an ionized gas rotated in the same direction as the younger counter-rotating stellar component~\citep[see examples in][]{ 2011MNRAS.412L.113C,2015A&A...581A..65C,2013MNRAS.428.1296J,2018A&A...616A..22P,2020MNRAS.495.1433Y}. Therefore, a more favourable explanation of the origin of galaxies with stellar counterrotation is a retrograde gas accretion~\citep{1998ApJ...506...93T,2014MNRAS.437.3596A,2017MNRAS.471L..87O} or mergers with gas-rich galaxies~\citep{1997ApJ...479..702T,2000MNRAS.316..315B,2008A&A...477..437D,2018MNRAS.481.3534S}.

\begin{figure}
\includegraphics[width=1\hsize]{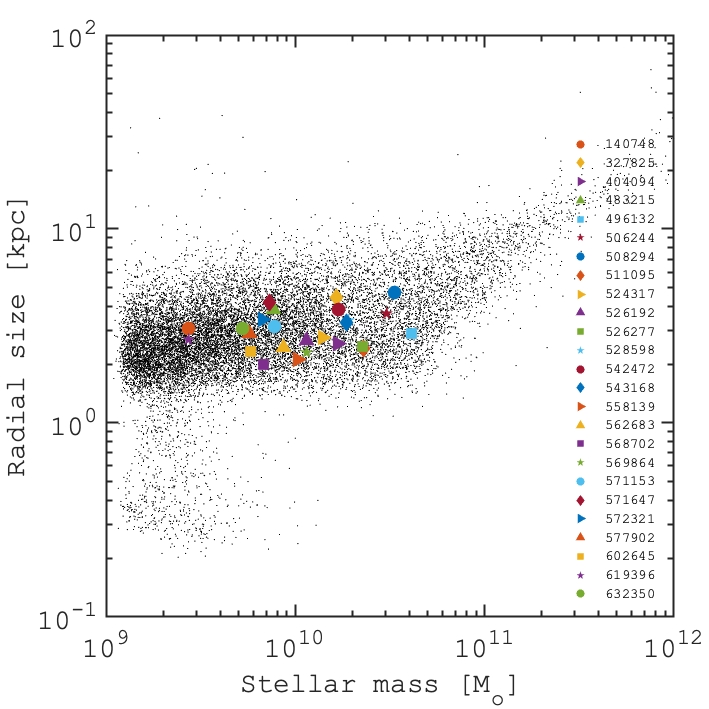}
\caption{Stellar mass - radial size relation for the IllustrisTNG100 galaxies. Black dots correspond to all the galaxies in the IllustrisTNG100 catalogue, while the coloured symbols show the galaxies with stellar counterrotation at $z=0$. The radial size here is calculated as the mean mass-weighted position of particles in the disc plane. In order to compare some kinematical characteristics of galaxies with counterrotation, for each galaxy in our sample we select 20 closest galaxies with similar masses and sizes based on this relation.}\label{fig::mass_size}
\end{figure}

Various simulations suggest that galaxy mergers are likely lead to the formation of an elliptical galaxy rather than of a regular thin disc galaxy~\citep[see, e.g.,][]{1993ApJ...403...74Q,1996ApJ...461...55T,2005A&A...437...69B,2011A&A...530A..10Q}. Therefore, a rather slow gas acquisition with retrograde spin from an external reservoir is the most probable formation scenario of the cospatial counterrotating component formation in galaxies. The best tool for studying a realistic gas infall onto galaxies is provided by cosmological simulations which significantly improved our understandig of the properties of simulated galaxies~(Illustris~\citep{2014MNRAS.444.1518V},  Horizon-AGN~\citep{2014MNRAS.444.1453D}, EAGLE~\citep{2015MNRAS.446..521S}, TNG~\citep{2018MNRAS.475..624N}) over the last decade. Some works based on cosmological simulations clearly demonstrate that, in a statistical sense, the simulated galaxies with misaligned components reproduce well many properties of observed galaxies~\cite{2020MNRAS.495.4542D, 2020ApJ...894..106K,2020arXiv200710342K}. Recently, by using different releases of the Illustris cosmological simulations, \cite{2019ApJ...878..143S} and \cite{2020MNRAS.492.1869D} made attempts to understand  mechanisms that lead to the formation of misaligned~(including counterrotating) components.  Although the authors used slightly different selection criteria, both works generally agree with each other that the formation of misaligned components correlates with the central black hole~(BH) activity, which could imply a generic link between the two phenomena. In particular, the BH feedback can remove pre-existing~(co-rotating) gas and make possible the accretion of retrograde gas onto the gas-poor galaxy. Alternatively, the retrograde gas accretion could increase the feeding of the BH and, thus, the AGN activity is the result of the counterrotating component formation. Therefore, the origin of the galaxies with counterrptation and its relation to various processes such as the gas accretion, interaction with other galaxies, stellar and AGN feedback and other effects, remains unclear.

\begin{figure*}
\includegraphics[width=1\hsize]{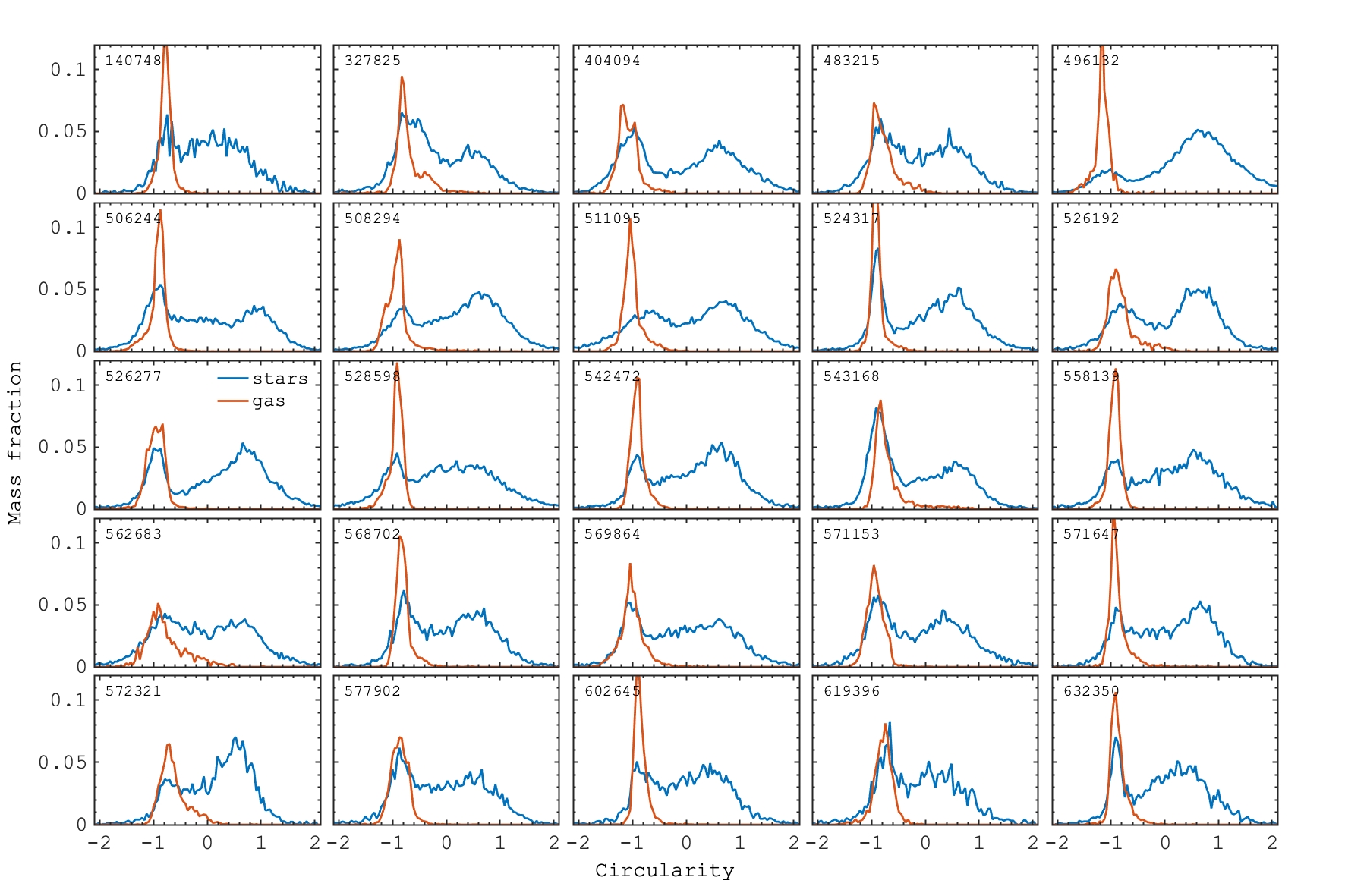}
\caption{Distribution of circularity for stars~(blue line) and gas~(red line) in the galaxies from our sample, at $z=0$. Although we make our selection by using the stellar counterrotation only, the gaseous component always rotates in the same direction as the retrograde stars providing an evidence for the evolution-based link between them.}\label{fig::circularity0}
\end{figure*}

In this work we aim to shed light on the formation paths and internal evolution of galaxies with extreme misalignment at $z=0$ wiht the help of the cosmological simulations TNG100 in the IllustrisTNG suite. We focus on a rather limited but still representative sample of galaxies with prominent stars-stars counterrotation. In particular, we investigate what causes the formation of the second~(counterrotating) stellar component and what kind of age-chemo-kinematic features can be used to constrain the formation paths in the real galaxies. We also provide some new models in support of results that we find in the Illustris simulations. The paper is structured as follows. In Section~\ref{sec::data} we describe the data and the criteria of the used galaxies selection. In Section~\ref{sec::global} we describe global characteristics and evolution of galaxies with counterrotation. In Section~\ref{sec::ages} we focus on the spatial scales, ages and metallicities of co- and counterrotating stellar components in galaxies. In Section~\ref{sec::origin} we explore the origin of the galaxies with counterrotation by making an in-deep analysis of the structure and mass evolution and gas kinematics. In Section~\ref{sec::SPH} we provide new $N$-body/hydrodynamical models of the formation of the counterrotating galaxies in order to demonstrate in detail the gas replacement due to a retrograde gas infall and its connection with the feeding of central regions in the galaxies. Finally, in Section~\ref{sec::conclusion} we summarize the main results of our work.

\section{Data and sample selection}\label{sec::data}
In this work we use the publicly available data from the IllustrisTNG cosmological magnetohydrodynamical simulations~\cite{2018MNRAS.473.4077P,2018MNRAS.475..624N, 2018MNRAS.475..676S, 2018MNRAS.480.5113M, 2018MNRAS.477.1206N,2019ComAC...6....2N}. For the purpose of this study we use the sublink merger trees~\citep{2015MNRAS.449...49R} and select the galaxies in the IllustrisTNG100 simulations from the redshift of $z \approx 5$ to the present epoch, $z = 0$. These simulations successfully reproduce the scaling relations and evolution of galaxy sizes~\citep{2018MNRAS.474.3976G}, including the formation of realistic disc galaxies~\citep{2019MNRAS.490.3196P}, the gas-phase mass-metallicity relation~\citep{2019MNRAS.484.5587T}, and the diversity of kinematic properties observed in the MW-type galaxies~\citep{2018MNRAS.481.1950L} and ETG halos~\citep{2020arXiv200901823P,2020A&A...641A..60P}. The Illustris simulations have been used to investigate various processes of galactic evolution, including the gas-stripping phenomena~\citep{2019MNRAS.483.1042Y,2020A&A...638A.133L}, central black hole activity~\citep{2019MNRAS.484.4413H,2020arXiv200800005D}, star formation quenching~\citep{2018MNRAS.474.3976G,2020MNRAS.491.4462D}, and many others. These results ensure that the investigation of properties of different types of galaxies from the Illustris cosmological simulations brings us close to understanding the nature of real objects including the galaxies with counterrotation.

From the snapshot corresponding to $z = 0$, we select subhalos identified with the Subfind algorithm~\citep{2001MNRAS.328..726S} and galaxies with a significant fraction of counterrotating stars. Following ~\cite{2003ApJ...591..499A}, we calculate the circularity parameter $\rm \epsilon = V_{rot} / V_{circ}$, where $\rm V_{rot}$ is the rotational velocity of star~(or gas) particle and $\rm V_{circ}$ is the circular velocity value. We assume that positive values of $\epsilon$ correspond to the main~(or hot) galaxy componentsm while negative $\epsilon$ implies the rotation in the opposite direction~(or counterrotation). Our initial sample  contains galaxies with a rather cold disc~(at least $30\%$ of stars have $\epsilon>0.5$) and at least $30\%$ of stars rotate in the opposite direction~($\epsilon<-0.5$). We exclude spuriously identified galaxies and galaxies with a dominant non-rotating component~(i.e. a  massive central spheroid with a broad distribution of $\epsilon$ around 0). In the end, we visually identify galaxies with two distinct peaks of the $\epsilon$ distribution near $\pm 1$. The final sample that we study consists of $25$~galaxies. 

It should be noted that only the stellar circularity has been used to identify the galaxies with counterrotation, without looking at the gas kinematics. We trace the present day structure and kinematics together with the assembly history of these galaxies, up to a lookback time of $12$ Gyr. Throughout the paper we use the host component notation to refer to the stellar populations that formed first. We orientate the galaxies in the way that the host components have dominant positive circularity. Therefore the stellar (and gaseous) components with negative circularity are referred to as counterrotating.

\begin{figure*}
\includegraphics[width=1\hsize]{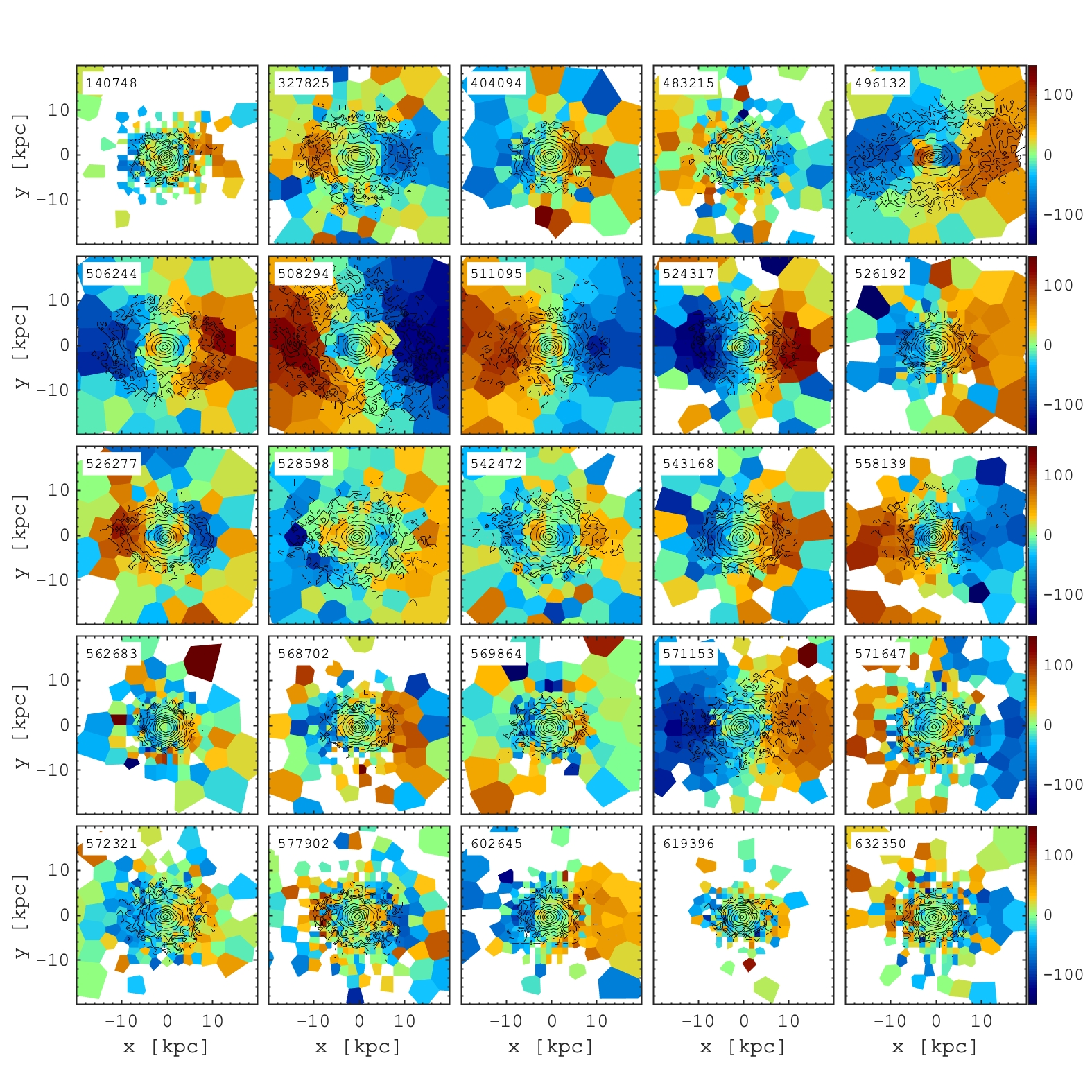}
\caption{The sample of galaxies with prominent stellar counterrotation at $z=0$. The maps show the line-of-sight stellar velocity distributions where all galaxies are inclined by $45^\circ$ with respect to the disc plane. The black contours depict the stellar density distribution. The presence of the counterrotation is seen well in some of the galaxies as a change of the rotation direction between the inner and outer regions. In each frame the number corresponds to the subhalo-id in the IllustrisTNG catalogue. The colour bars depict the line-of-side velocity component in \kmps.}\label{fig::vel_los}
\end{figure*}

First, we give a brief overview of the global characteristics of the sample of galaxies with counterrotation and compare them to the entire IllustrisTNG galactic population. In Fig.~\ref{fig::mass_size} we plot the stellar mass -- radial size relation for all the galaxies in TNG100, where the radial size is calculated as the mean mass-weighted position of stellar particles along the galactocentric cylindrical distance. Although our definition of the galaxy size differs from e.g. the stellar half-mass size, our mass -- size relation is very similar to the one presented by ~\cite{2018MNRAS.480L..18Z}. In Fig.~\ref{fig::mass_size} we also highlight all the galaxies with stellar counterrotation at $z=0$ by colour marks. In terms of stellar masses and sizes, the galaxies with counterrotation seem to be typical objects among the TNG100 galaxies. However, we do not find any galaxies with counterrotation with stellar mass larger than $\approx 3\times 10^{10}$~\Msun. Since the decoupled kinematics is a clear evidence for the external origin of one of the components, the upper mass limit for counterrotating galaxies would impose some constrains on their formation mechanisms. Perhaps, in the case of galaxies more massive than $\approx 3\times 10^{10}$~\Msun, one needs higher masses of accreted gas to form a counterrotating component, while the accretion of such a large amount of gas could be rare and hence is not observable among the IllustrisTNG galaxies.

\begin{figure*}
\includegraphics[width=1\hsize]{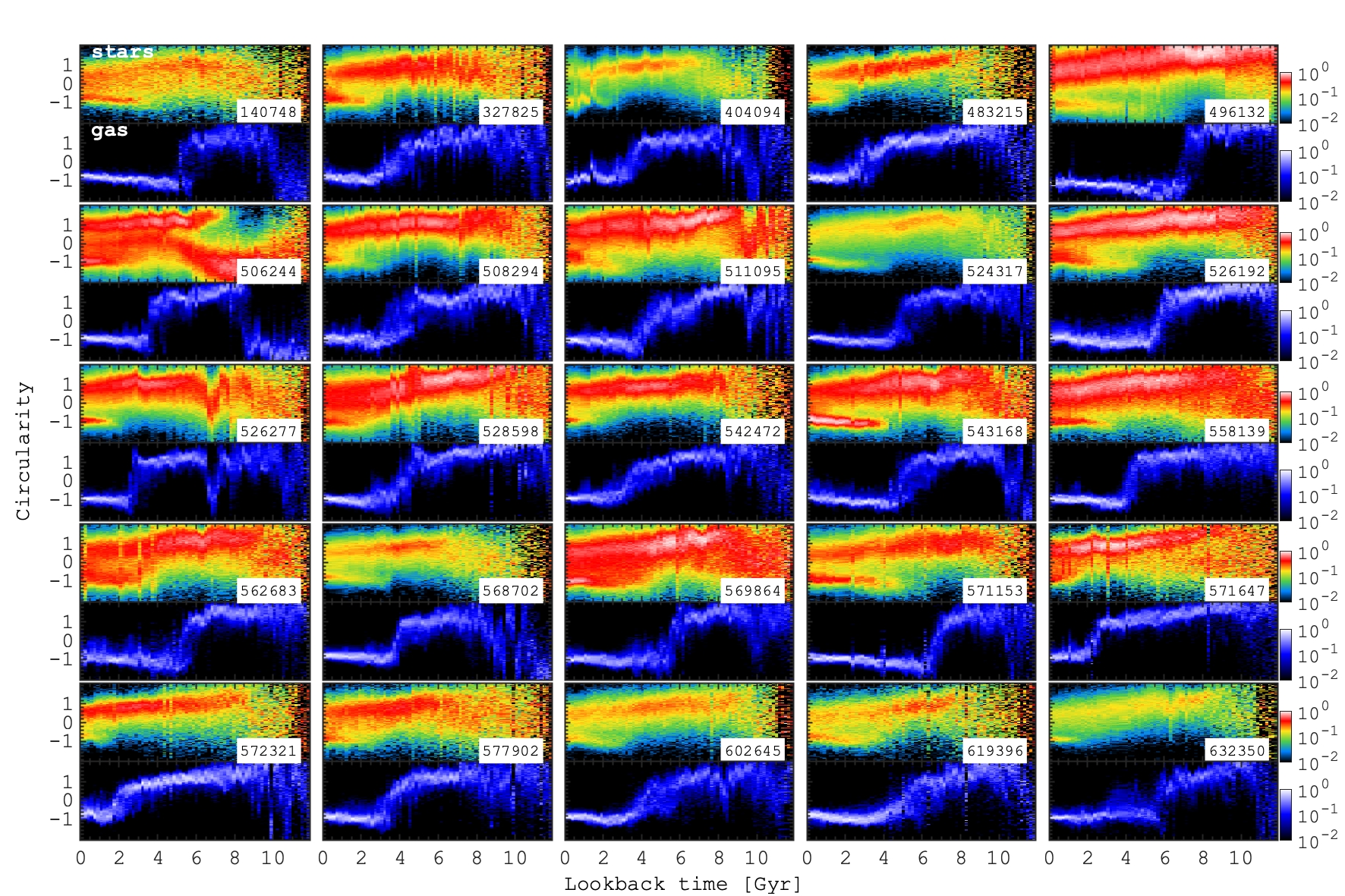}
\caption{Each frame shows the evolution of the circularity distribution for stars~(top subframes) and gas~(bottom subframes) in a galaxy with counterrotation. The distributions are given in the log-scale and are normalized by the total stellar (or gas) mass at a given snapshot. The host component evolution can be traced along the overdensities with a positive circularity, while counterrotating components seen well as the overdensities with negative circularity. The exceptional galaxies are 140748, 506244 and 543168  where three components can be found in the circularity distribution of stars. Note that all the galaxies in our sample contain both stellar and gaseous components rotating in the opposite direction with respect to the host stellar component. Obviously, the gaseous components always rotate in one direction where the beginning of the counterrotating component formation can be identified at the moment of the gas rotation direction. For  most of the galaxies the stellar counterrotation starts to appear after the counterrotating component settles down near the circularity $\epsilon \approx -1$. The colour bars depict the mass fraction normalized by the maximum value at a given time.}\label{fig::circularity_evolution}
\end{figure*}

\section{Stellar and gaseous counterrotation}\label{sec::global}

Next we consider the global kinematic properties of the sample of galaxies with counterrotation. In Fig.~\ref{fig::circularity0} we present the distributions of both stellar and gaseous circularity at $z=0$. Although our sample selection has been performed using the stellar circularity only, we surprisingly find that all galaxies in our sample also contain a prominent gaseous counterrotation, which already suggests some genetic connection between the gaseous component and the counterrotating stars. This link is also supported by the fact that the gas rotation is similar to the colder stellar component, implying that these dynamically colder stars were likely formed more recently from the retrograde gaseous disc. 

In Fig.~\ref{fig::vel_los} we present the line-of-sight stellar velocity maps adopting an inclination angle of $45^\circ$ for all the galaxies. We can identify prominent counterrotating components in different parts of the galaxies. In particular, about half of our sample demonstrates a clear change of the direction of stellar rotation between the inner and outer parts. This suggests that the host and counterrotating components dominate in different disc regions. Such a picture is quite similar to one observed in NGC~448 where a change of the direction of the rotation from the inner to the outer regions of the galaxy is observed due to the presence of an extended disc component that rotates in a noticeably different manner with respect to that of the main stellar body~\citep{2019A&A...623A..87N}. 

For the further analysis of the counterrotating galaxies we demonstrate the evolution of the circularity distribution for the stellar and gaseous components over the last $12$~Gyr. Fig.~\ref{fig::circularity_evolution} shows the distribution of the mass of stars~(top subframes) and gas~(bottom subframes) in the circularity-lookback time diagram. At each snapshot we normalize the distributions by the maximum value. The evolution of the host stellar component circularity~($\epsilon >0$) has a declining with time pattern, implying a gradual decrease of the number of stars on nearly circular orbits~(near $\epsilon \sim 1$) due to a gradual heating of the stellar disc. 

The evolution of the gas circularity is more diverse. At early times the rotation of the gas is aligned with the host stellar population which is likely to be made of this cold gas. However, in all the galaxies at different times, we observe a sharp change in the direction of rotation of the gas where the maximum of the distribution crosses the value $\epsilon=0$. This pattern is visible across the entire sample and suggests a similarity in its formation history. Once the gas settles down around the negative circularity $\epsilon \sim -1$, new populations of stars start to appear as the second sequence in the stellar distribution of the circularity~($\epsilon < 0$). Therefore, the stellar circularity distribution shows a clear bimodality over the last Gyr.

\begin{figure*}
\includegraphics[width=1\hsize]{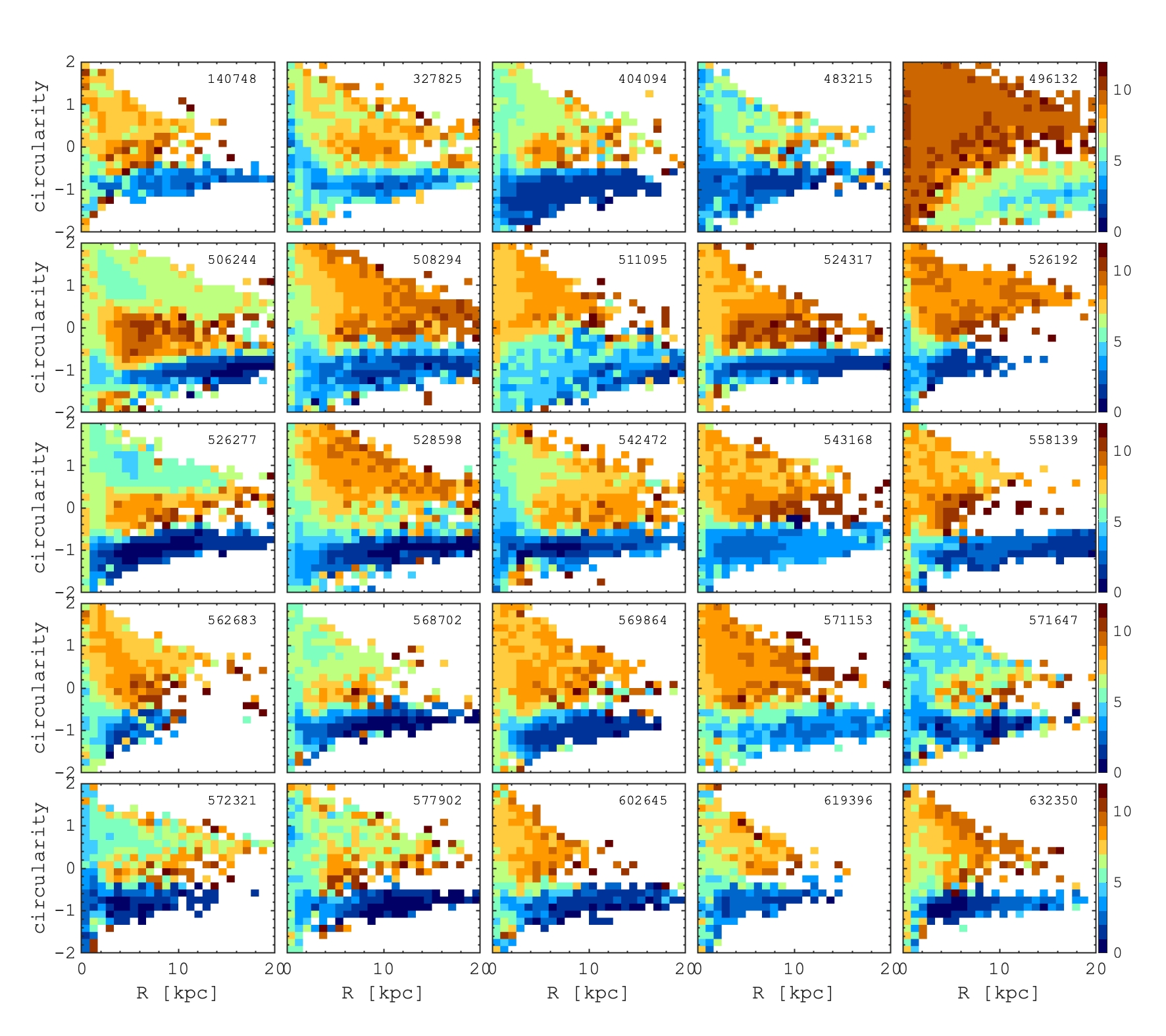}
\caption{The circularity -- galactocentric distance relation at $z=0$ for the galaxies with counterrotation. The maps are colour-coded by the mean stellar age in Gyr. Three clear populations can be identified: an oldest non-rotating components near the $\epsilon = 0$, a mid-age old dynamically hot, discy-components~($\epsilon>0$), and a young and dynamically cold counterrotating discs. Significant radial extension of both disc components illustrates that all galaxies in our sample show the cases with a large-scale counterrotating stellar component with substantial spatial overlap.}\label{fig::circularity_radius}
\end{figure*}

\begin{figure*}
\includegraphics[width=1\hsize]{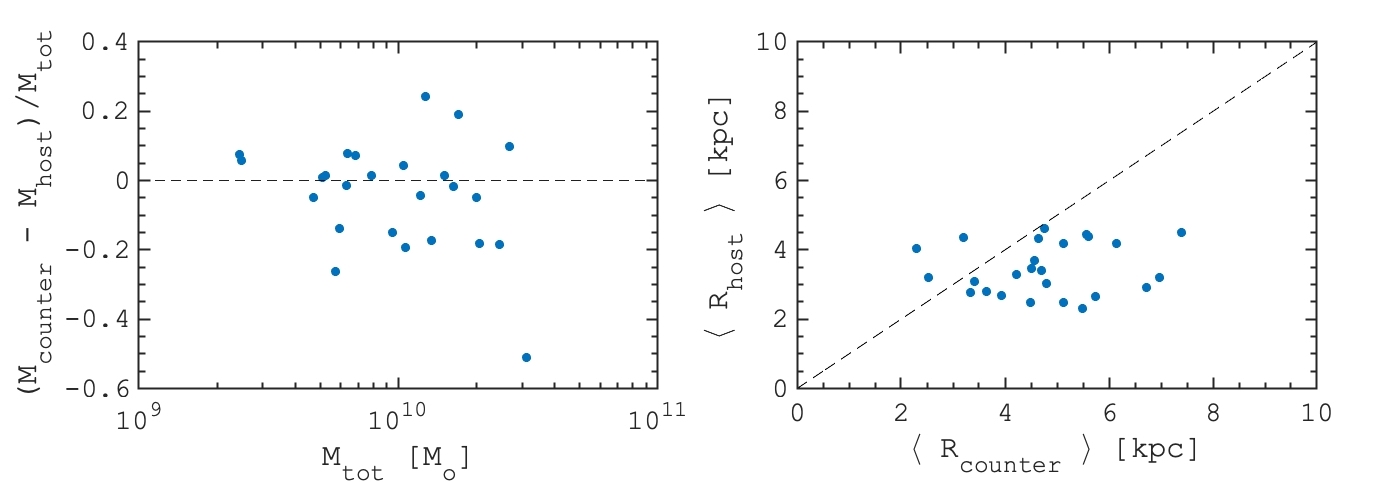}
\caption{ {\it Left:} a relative mass fraction of the counterrotating component where the $\rm M_{counter}$ indicates the total mass of counterrotating stars, $\rm M_{host}$ is the mass of stars in the host component and $\rm M_{tot} = M_{host} + M_{counter}$. {\it Right:} a relation between the radial size of the counterrotating~($\rm R_{counter}$) and host~($\rm R_{host}$) components. For each component both values are calculated as the mean mass-weighted galactocentric distance of star particles.}\label{fig::mass_size_compare}
\end{figure*}

\begin{figure}
\includegraphics[width=1\hsize]{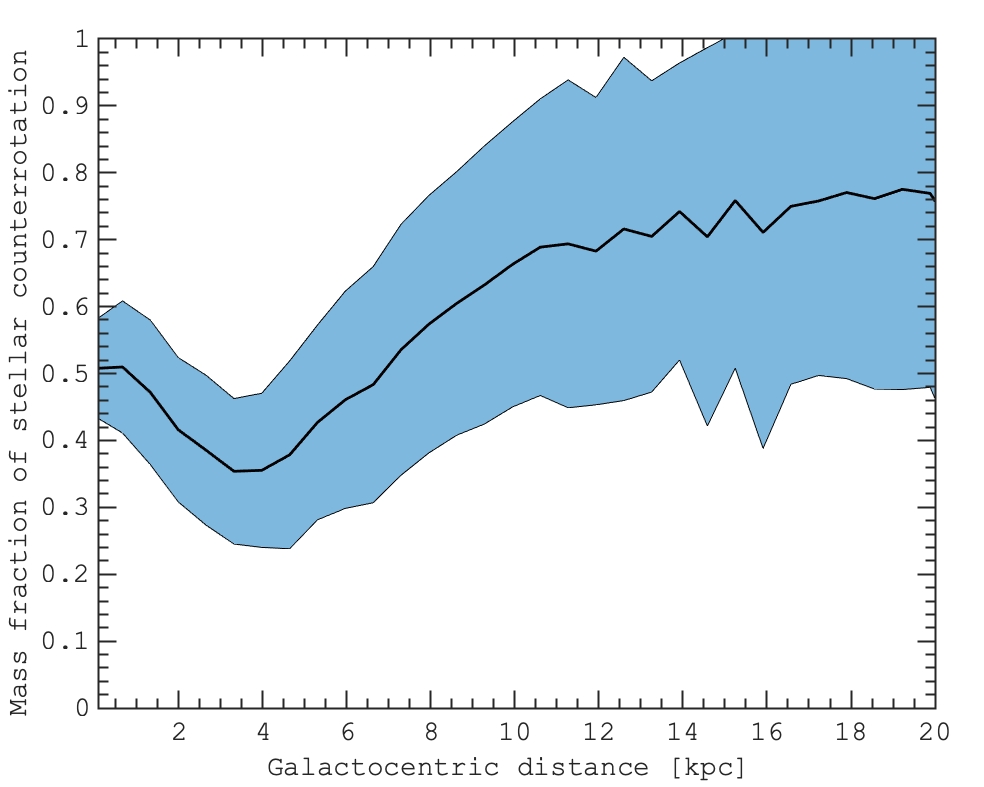}
\caption{The radial distribution of stellar counterrotating component mass fraction at $z=0$. The profile represents the averaged distribution among all galaxies in our sample where the scatter represents the standard deviation and is shown by the blue area.}\label{fig::mass_fraction}
\end{figure}

There are a few galaxies (140748, 506244 and 543168) that experience a more remarkable evolution~(see Fig.~\ref{fig::circularity_evolution}). Apparently, the gas in these galaxies changed the direction of its rotation twice. As a result, we observe the appearance of three distinct stellar components. The first one was formed at early times, and after the first flip of the gas rotation it was dragged into a dispersion-supported component with the circularity of $\epsilon \sim 0$. The second stellar component is likely formed from the gas with $\epsilon<0$, and the last~(third) stellar component was formed after the gas rotation flipped once again. Therefore, the galaxies end up with having two counterrotating stellar discs and a bulge-like component as well. These galaxies show a similar evolution to the simulated galaxies presented by~\cite{2014MNRAS.437.3596A} which experienced two episodes of accretion from two filaments with different angular momentum direction.

Next, in Fig.~\ref{fig::circularity_radius} we present the distribution of the circularity parameter for stars as a function of galactocentric distance, colour-coded by the mean stellar age. As we already found, the host components appear to be dynamically hotter and, at the same time, radially-shorter in comparison than the colder and more radially extended counterrotating stellar components. There is a clear age  segregation in the majority of the galactic components. In particular, non-rotating stars with $\varepsilon\approx 0$ represent the oldest population. Then the host disc components~($\varepsilon\approx 1$) are substantially younger, while the counterrotating stars form the youngest population of the galaxies. We find a rather smooth age-transition between the components, suggesting a continuous evolution of our galaxies that leads to the formation of these kinematically distinct components.

\section{Properties of counterrotating components}\label{sec::ages}

Before moving forward in understanding the origin of the counterrotation~(both stellar and gaseous), we provide a more detailed description of the global properties of the components of the galaxies we analyze. 

In Fig.~\ref{fig::mass_size_compare}~(left) we compare the stellar masses of the host and counterrotating components. The relative mass of each component does not correlate with the galactic total mass. The component masses in the simulated galaxies are  comparable with observed galaxies   ~(e.g. $\approx 30\%$ in NGC 448~\citep{2016MNRAS.461.2068K}, $\approx 45\%$ in NGC~1366~\citep{2017A&A...600A..76M}). In some galaxies the counterrotating component could be even more massive than the host one. Comparison of the radial scales of the counterrotating and host components is shown in Fig.~\ref{fig::mass_size_compare}~(right). In the majority of galaxies the counterrotating component is at least twice larger than the host one. To understand better the radial structure of the counterrotating components in Fig.~\ref{fig::mass_fraction}, we present the mass fraction of counterrotating stars as a function of radius averaged across all galaxies in our sample. In the innermost galaxy, up to the host component's effective size of~($2-4$~kpc), the fraction of counterrotating stars is nearly constant and contributes about a half of the mass. At larger radii the mass of counterrotation gradually increases and can reach up to $100$\%. Therefore, we can conclude that the stellar counterrotating components in our sample are presented by radially extended discs surrounding the compact host ~(see also Fig.~\ref{fig::circularity_radius}). From real observations we can point at e.g. an extreme case with the extended counterrotation in UGC~1382, which possesses a giant counterrotating gaseous disc with the radius of $80$~kpc~\citep[][Saburova et al. in prep.]{saburova2020}. At the same time, in some galaxies with counterrotation (e.g. IC~719 \citep{2013ApJ...769..105K} and NGC~448, \citep{2016MNRAS.461.2068K}), the younger counterrotating stellar component is more concentrated to the centre than the host one.

\begin{figure*}
\includegraphics[width=1\hsize]{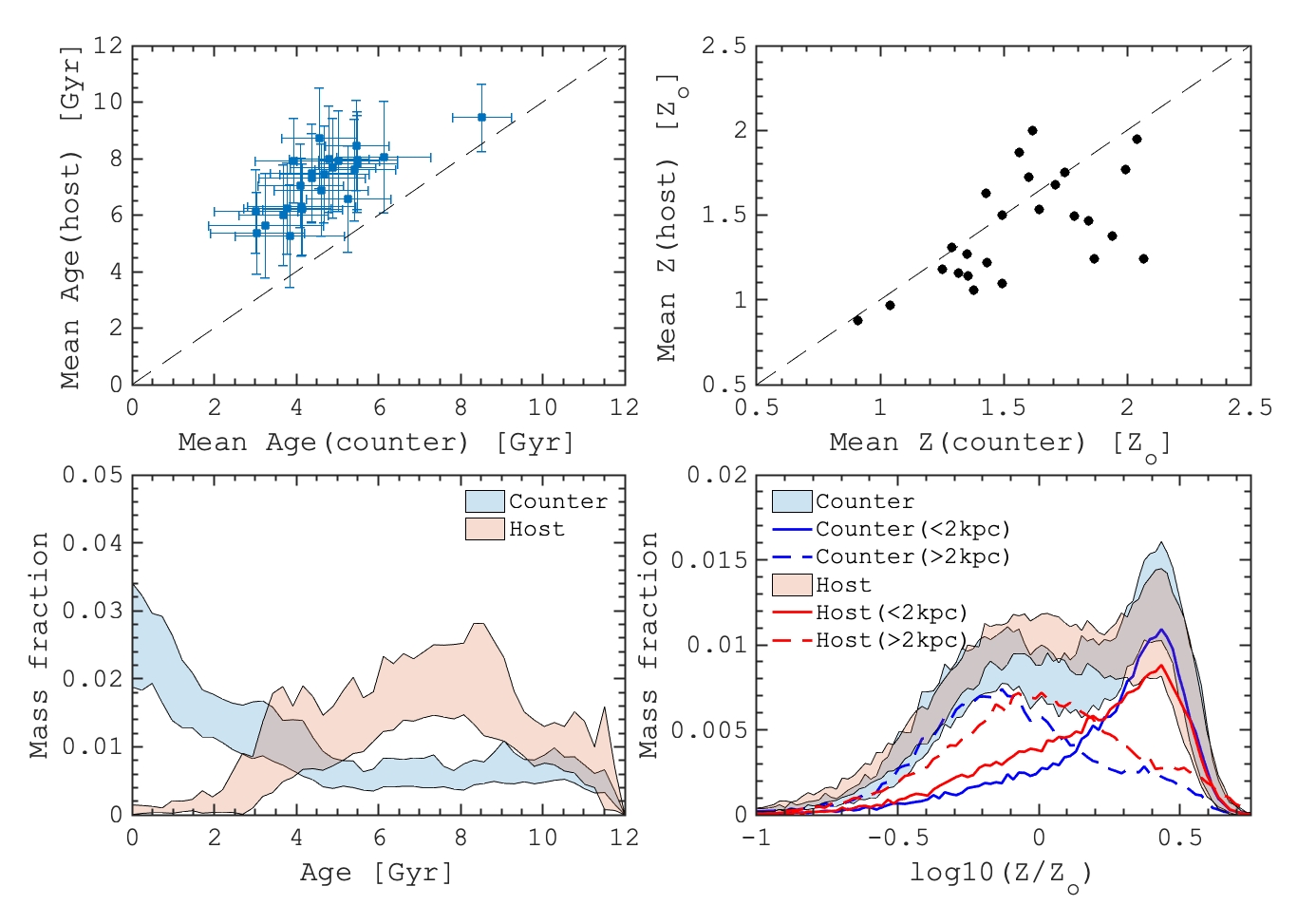}
\caption{ {\it Top left:} the mean age of counterrotating stars versus the mean age of the host component. The errorbars correspond to the standard deviation of the age measurements of both components. {\it Top right:} the mean metallicity of counterrotating stars versus the mean metallicity of the host component. {\it Bottom left:} the mean age distribution of stars in counterrotating~(blue) and host~(red) components averaged among the entire sample of 25 galaxies. The vertical scatter depicts the standard deviation. {\it Bottom left:} metallicity distributions for both components averaged among the entire sample of 25 galaxies: the blue and red colours correspond to counterrotating and the host components, respectively. The vertical scatter depicts the standard deviation.}\label{fig::age_met}
\end{figure*}

In Fig.~\ref{fig::age_met}~(top left) we present the relation between the mean ages of stars of the host and the counterrotating components. The striking feature is that in all the galaxies the counterrotating stars are significantly younger with respect to the host component stars, which is expected because the source of the most recent star formation is the counterrotating gas. Such a relation is in favour of the formation of younger counterrotating stellar components from externally acquired gas, and not from infalling dwarfs or major merger. The age difference of about $1-3$~Gyr that we find is significant and agrees well with observations of galaxies with counterrotation where the younger stellar component was found to be co-rotating with the gas~\citep[see, e.g.,][]{2011MNRAS.412L.113C, 2015A&A...581A..65C}, which supports the gas accretion scenario. A large scatter of ages in both components inspires us to test the star formation histories of both components~(see Fig.~\ref{fig::age_met}, bottom left). The host component stars have been formed mainly between $4$ and $10$ Gyr ago with a peak at around $8-9$~Gyr ago, and a negligible mass fraction of the host component is formed over the last $3$~Gyr. In the contrary, the counterrotating components show a steep rise of the stellar mass during the last few Gyrs.

The different star formation histories likely result in a substantial difference in the chemical abundances of the host and counterrotating stars. In Fig.~\ref{fig::age_met}~(top right) we compare the mean metallicity of these two populations. The metallicity is derived as the total mass of all metals~(elements above He) in solar units. A more recent and intense star-formation turns out in the higher mean metallicity of counterrotating components. The difference seen for most of the galaxies is still remarkable. This result is somewhat counter-intuitive because the metallicity of the retrograde gas accreted from filaments should be close to primordial. On the other hand, this external gas mixed with a pre-existing one. The expected dilution of metals can be vanished by the intense star formation in the center. At the same time, it can be still found at the outskirts where the star formation is usually more quiescent. Indeed, the outer parts of counterrotating components show lower stellar metallicities in the comparison to that of the hosts (see Fig.~\ref{fig::age_met}, bottom right). This can indicate that the outer counterrotating stars were formed from the more metal-poor gas than it was before the accretion, as it is expected for the accretion from filaments.

\begin{figure*}
\includegraphics[width=1\hsize]{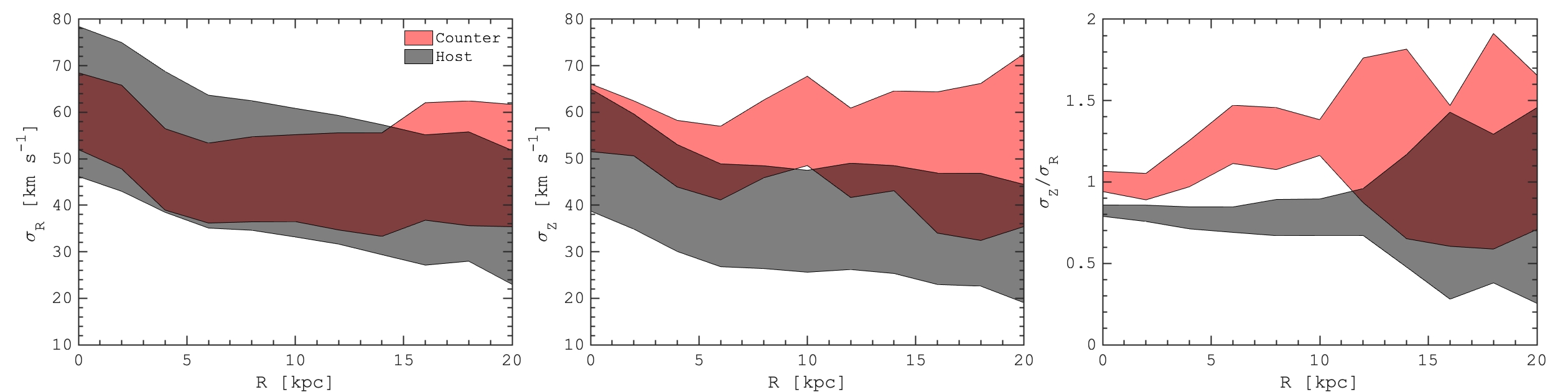}
\caption{Mean radial~(left, $\rm \sigma_R$), vertical~(center, $\rm \sigma_Z$) and vertical-to-radial velocity dispersion ~(right, $\rm \sigma_Z/\sigma_R$) profiles averaged for the galaxies with counterrotation~(red) and for the reference sample of galaxies without counterrotation~(gray). The vertical scatter depicts the standard deviation. The radial velocity dispersion profiles in counterrotating galaxies are similar to our reference sample, being slightly flattened, while the vertical velocity dispersion is higher, which results to a significant difference in the $\rm \sigma_Z/\sigma_R$ distribution.}\label{fig::mean_dispersion}
\end{figure*}

Another question we wish to explore in this section is the difference between the internal kinematics of components of the galaxies with counterrotation which we compare to the sample of typical galaxies selected according to the same size and mass~(see Fig.~\ref{fig::mass_size}). In Fig.~\ref{fig::mean_dispersion} we show the radial profiles of the averaged~(among all the galaxies) radial~($\sigma_R$), vertical~($\sigma_Z$) velocity dispersion and vertical-to-radial velocity dispersion ratio~($\sigma_Z/\sigma_R$). The galaxies with counterrotation are shown in red, the normal galaxies are designated by gray. The radial velocity dispersion profiles seem to be similar for the normal and counterrotating galaxies, while the vertical velocity dispersion is slightly higher. Once we compare the velocity dispersion ratio, we find that normal galaxies show a flat radial profile with $\sigma_Z/\sigma_R\approx0.8$, in agreement with some empirical studies suggesting its monotonic variation with the Hubble type. In particular, it is found that $\sigma_Z/\sigma_R\approx0.8$ in Sa-type galaxies, and it decreases down to $\approx0.2$ in Scd galaxies~\citep{2003AJ....126.2707S, 2012MNRAS.423.2726G}. Meanwhile, the counterrotating galaxies have $\sigma_Z/\sigma_R$ ratio larger than unity across the whole disc. Such a behaviour has been predicted by \cite{2017A&A...597A.103K}, where the strong vertical heating due to the bending instabilities leads to a significant increasing of the vertical-to-radial velocity dispersion ratio on a short dynamic time-scale. 

The vertical disc heating points at the importance of the counterrotating components for shaping the velocity dispersion profiles in general. Even though the studied galaxies represent an extreme case of the counterrotation, the presence of a small counterrotating components could be more obvious. Therefore, a small fraction of counterrotation in the disc could  provide a new channel of the formation of kinematically hot and geometrically thick stellar components. In particular, our results suggest that the formation of thick stellar discs can be partially provoked by the accretion of counterrotating components at early epochs. For example, it is now acknowledged that the Milky Way's main progenitor has accreted a few massive satellites about $9$ Gyr ago~\citep{2018ApJ...863..113H,2018Natur.563...85H}, and at least one of them, the Sequoia~\citep{2019MNRAS.488.1235M}, is found to be couterrotating. Since the epoch of the accretion events corresponds to the formation of chemically distinctive thick disc in the Milky Way, we suggest that it could be, at least partially, shaped by the accretion of counterrotating stellar component.

\section{Formation of the counterrotation}\label{sec::origin}
\subsection{Star-gas galaxies structure}
\begin{figure*}
\includegraphics[width=1\hsize]{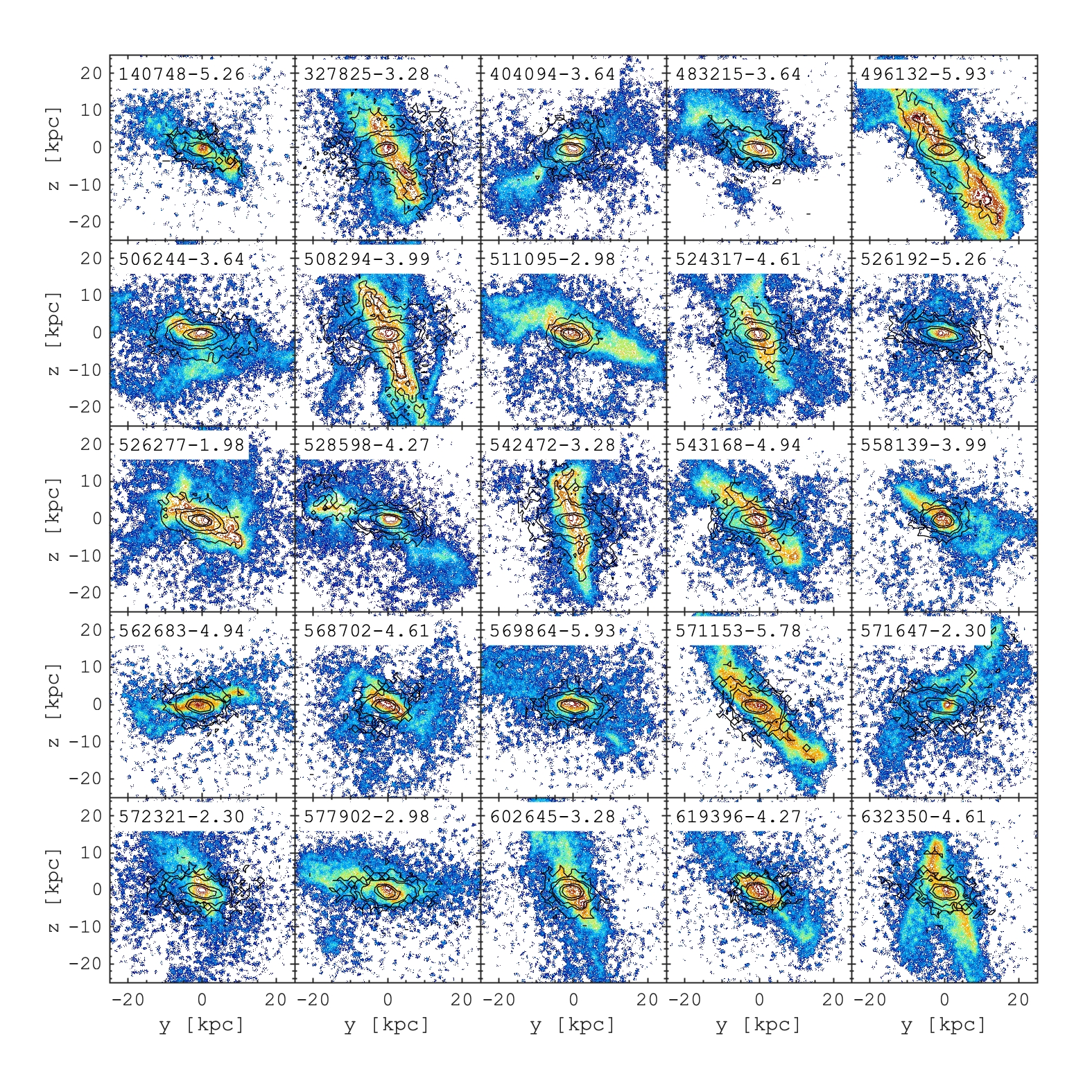}
\caption{The structure of stellar~(black contours) and gaseous~(colourmap) components close to the beginning of the gas accretion. In each frame the numbers correspond to the subhalo-id and lookback time. We orientated all galaxies to minimize the projected width of the infalling gaseous component. Some objects can be classified as polar ring galaxies.}\label{fig::PRGs}
\end{figure*}

\begin{figure*}
\includegraphics[width=1\hsize]{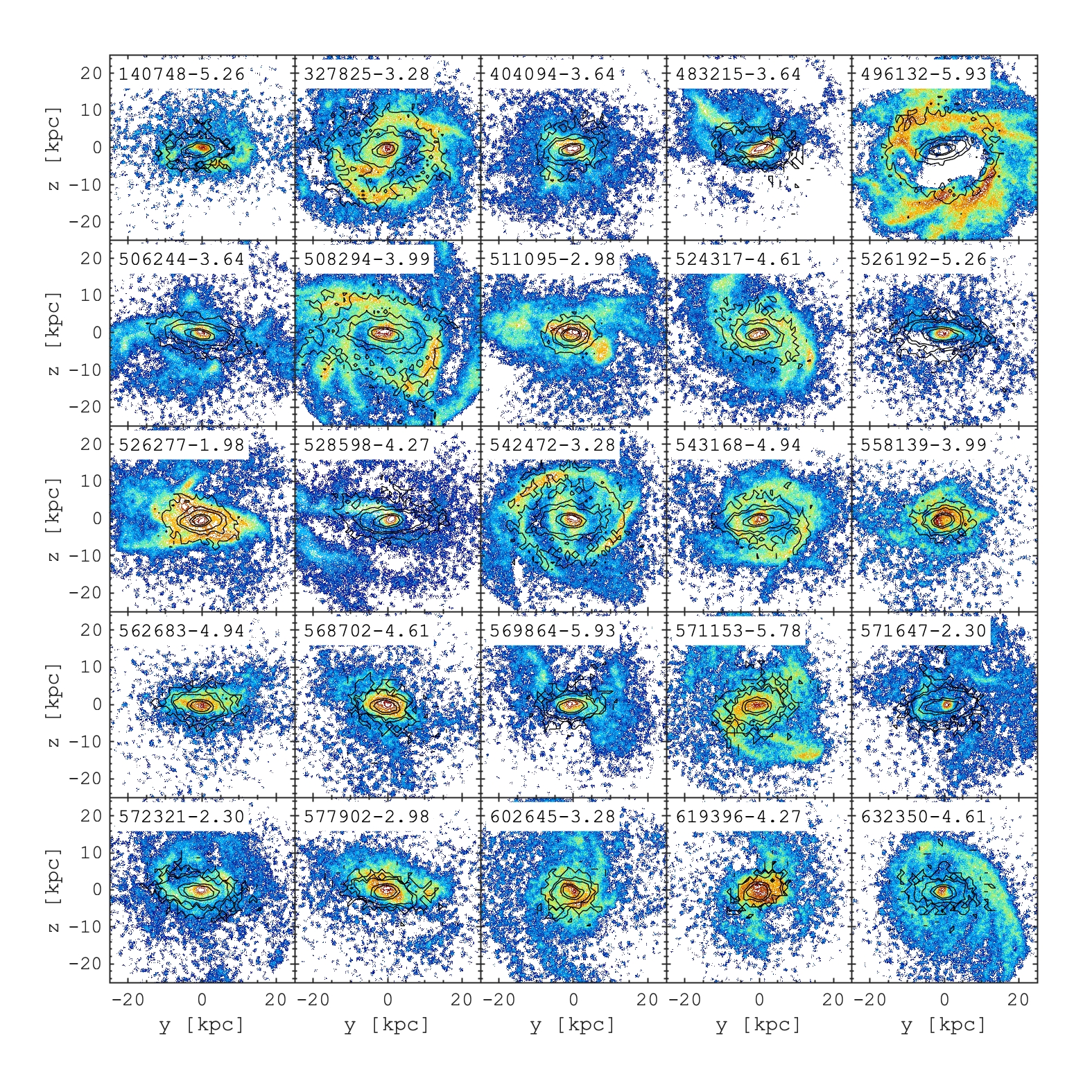}
\caption{The structure of stellar~(black contours) and gaseous~(colourmap) components close to the beginning of the gas accretion. In each frame the numbers correspond to the subhalo-id and the lookback time. We orientated all galaxies to maximize the projected width of the infalling gaseous component. Thereby the gaseous component can be seen as extended disc or ring with a spiral structure.}\label{fig::rings}
\end{figure*}

\begin{figure}
\includegraphics[width=1\hsize]{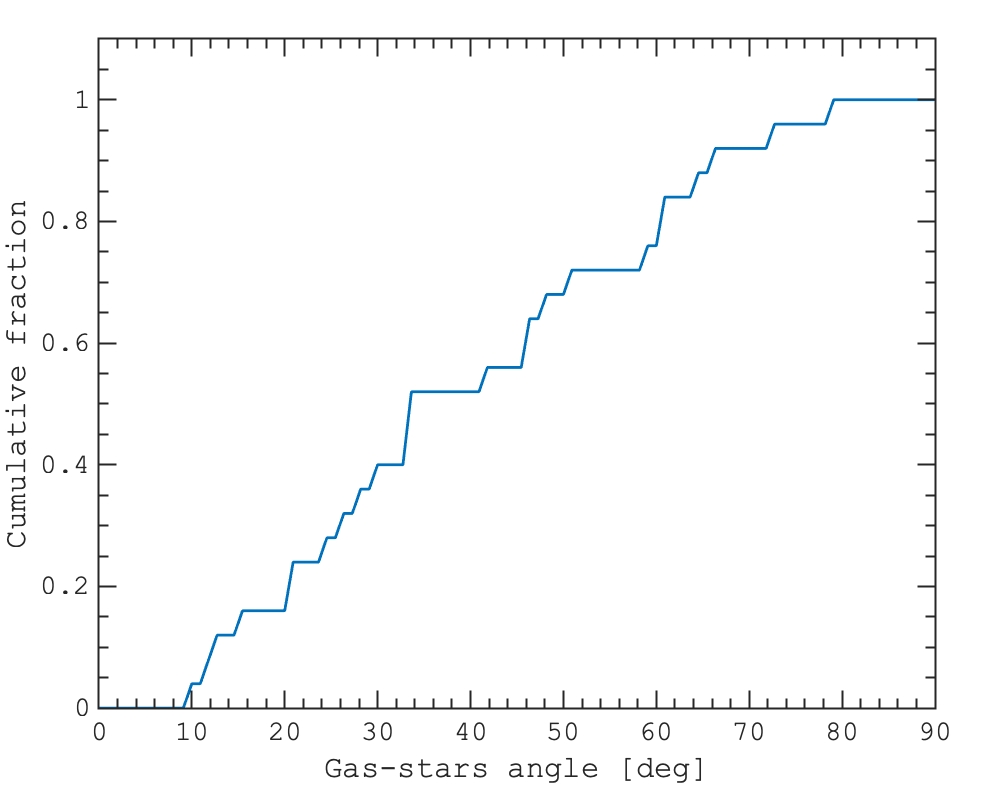}
\caption{Cumulative distribution of the relative angle between the stellar host component and off-plane gas measured near its infall episode. The distribution suggests that there is no preferable direction of the gas infall, as it is also well seen in Fig.~\ref{fig::PRGs}.}\label{fig::infall_angle}
\end{figure}

\begin{figure*}
\includegraphics[width=1\hsize]{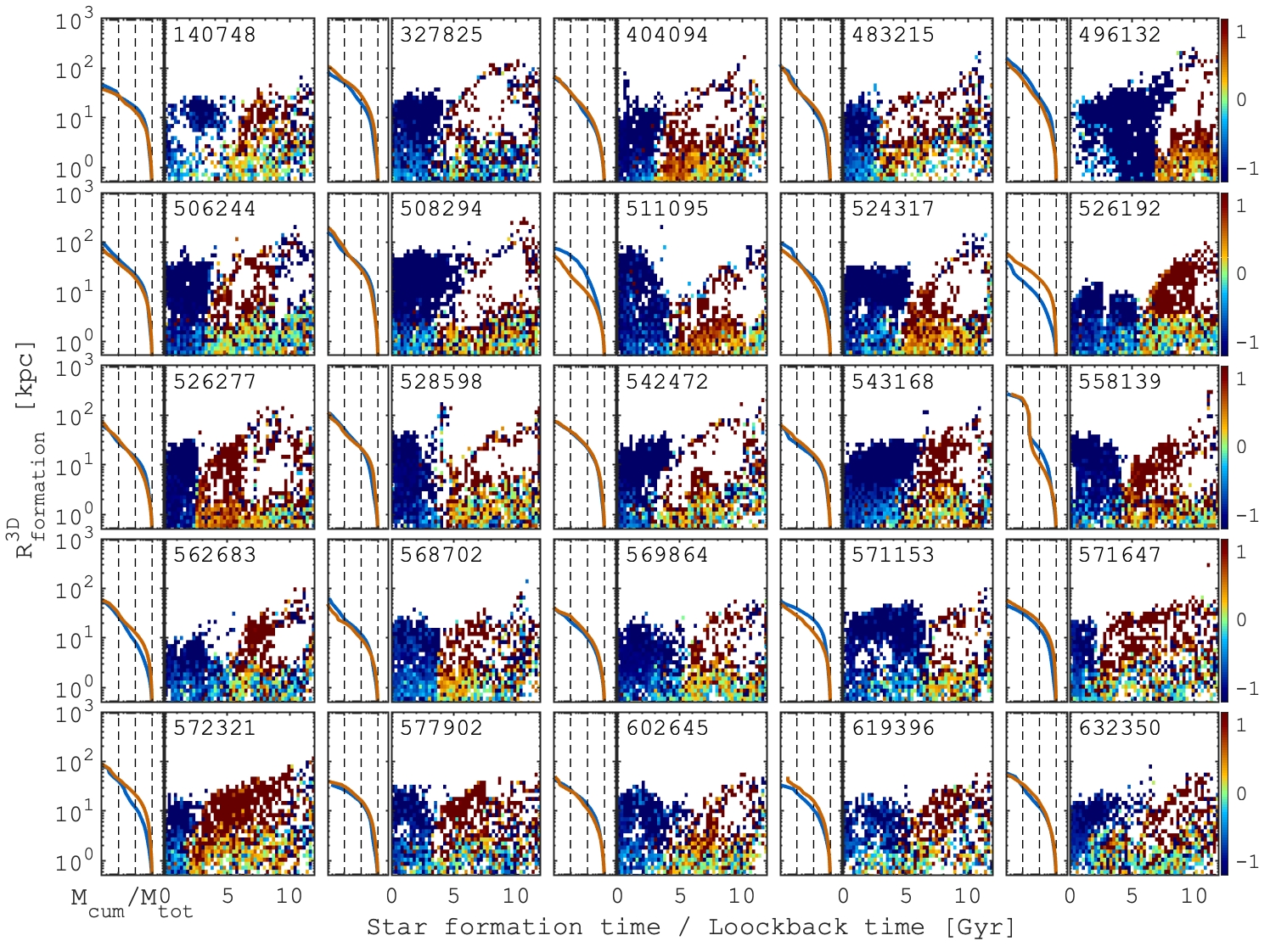}
\caption{Formation radius~($\rm R^{3D}_{formation}$, see Eq.~\ref{eq::formation_radius}) of stars as a function of the formation time colour-coded by the mean circularity value calculated at the end of the simulation. Subframes on the left show a cumulative mass distribution at the end of the simulation as a function of $\rm R^{3D}_{formation}$ for counterrotating~(blue, $\varepsilon<0$) and co-rotating stellar components~(red, $\varepsilon>0$) shown separately. The cumulative mass scale~($\rm M_{cum}/M_{tot}$) is represented by the vertical dashed lines highlighting the $0.01$, $0.1$ and $1$ levels, left to right.}\label{fig::formation_radius_time_circularity}
\end{figure*}

In this section we take a deep look into the formation history of the galaxies with counterrotation. First, we analyse the spatial structure of stellar and gaseous components of the galaxies close to the moment of the gas rotation reversal. We identified these moments for each galaxy individually as the time when the gas circularity distribution peaks at $\approx 0$~(see Fig.~\ref{fig::circularity_evolution}). The moment of the gas rotation reversal ranges from $2$ to $8$~Gyr ago, with a peak near $4$~Gyr ago~(see Fig.~\ref{fig::accr_time} in Appendix). In Figs.~\ref{fig::PRGs} and ~\ref{fig::rings} we show the density structure of stars~(black contours) and gas~(coloured maps) at that time. We find that most of the gas appears to be inclined with respect to the host stellar component, suggesting an external origin of this gas. The three-dimensional structure of the gas is highlighted by adopting two specific projections: in Fig.~\ref{fig::PRGs} we orientate the galaxies to minimize the projected width of the gas, while in Fig.~\ref{fig::rings} we show the orientation perpendicular to the former. In Fig.~\ref{fig::PRGs} the gaseous components have almost isotropic distributions of their orientation angles. To quantify this in Fig.~\ref{fig::infall_angle}, we show a cumulative distribution of the relative angle between the main stellar body and the off-plane gas~(see Fig.~\ref{fig::PRGs}).

In Fig.~\ref{fig::PRGs} some galaxies have a nearly-perpendicular orientation of the gas with respect to the main stellar body, while having a ring- or disc-like morphology. At the moment of the gas rotation reversal most of the galaxies can be classified as galaxies with polar rings~\citep[PRGs, see, e.g.,][]{1990AJ....100.1489W,2011MNRAS.418..244M}. Such a picture provides the first evidence of genetic connection between two types of multi-spin galaxies~\citep{1994AJ....108..456R}. In particular, we find that some  of the counterrotating  galaxies could have a polar ring structure some Gyrs ago. Although it is suggested that polar structures can be stabilized for a long time by the (triaxial)~dark matter potential~\citep[see, e.g.,][]{1994AJ....107..958A,1994ApJ...436..629S,2012MNRAS.425.1967S,2014MNRAS.441.2650K}, in our galaxies these polar rings~(or discs) settle down into the plane of the host stellar component rather quickly, within $1-3$~Gyrs. Other signatures of instability of the gaseous structures are the prominent one- or two-armed spiral structures. Most likely, the asymmetric, one-arm structures represent some gas flows spiraling down to the stellar disc. Meanwhile, two-armed morphology can result from interactions with the host disc potential~\citep{2013MSAIS..25...51K,2015BaltA..24...76M} or with a non-axisymmetric dark matter distribution~\citep{2012ARep...56...16K,2013MNRAS.431.1230K}. It is worth to mention that the appearance of nearly polar gaseous structures is also predicted at some stages of evolution of Milky Way-like galaxies~\citep{2020arXiv200606011R,2020arXiv200606008A,2020arXiv200606012R}.

A remarkable feature which we previously noticed in the circularity evolution~(see Fig.~\ref{fig::circularity_evolution}) is a delay in the stellar counterrotating component formation after the gas settled down in the counterrotating configuration. It suggests that the stellar component has been mainly formed in-situ instead of having been accreted. This can be also captured in Figs.~\ref{fig::PRGs},~\ref{fig::rings}, where very tiny stellar structures can be found outside the main stellar body. We do not see a signature of substantial amount of stars associated with the gas orbiting around the galaxy. Such a picture is in favour of the gas acquisition from filaments where mergers do not play a substantial role. To explore this issue in Fig.~\ref{fig::formation_radius_time_circularity} we present the relation between the star formation radius and the star formation time~(or loockback time). For simplicity, we define the star formation radius as a 3D spherical galactocentric distance:
\begin{equation}
\Oo \rm  R^{3D}_{formation} = \sqrt{ (x_f - x_0)^2 + (y_f - y_0)^2 + (z_f - z_0)^2 }\,\label{eq::formation_radius}
\end{equation}
where $\rm x_f,y_f,z_f$ are the coordinates of newly formed stars and $\rm x_0,y_0,z_0$ are the coordinates of the galactic center at the time of the star formation. In Fig.~\ref{fig::formation_radius_time_circularity} the relation between the birth time and the formation radius is colour-coded by the circularity value derived at the end of the simulation. Subframes with the distributions show the cumulative mass fraction as a function of the formation radius for the host~(red) and counterrotating~(blue) stellar components. We find that the vast majority of the counterrotating~(at $z=0$) stars were formed in-situ. In particular, about $90\%$ of them were formed within $\approx10-20$~kpc from the galactic centre and only in a few galaxies there is a small fraction of retrograde stars formed at larger distances from the galactic centre. Note that some of the galaxies have clear gaps in the diagrams in Fig.~\ref{fig::formation_radius_time_circularity}, which correspond to the active star formation in the accreted retrograde gas at the galactic outskirts or in polar~(or inclined) gaseous structures. Such a picture could correspond to the star formation in misaligned~(or even counterrotating) ring-like structures found at the outskirts of some galaxies~\citep{2019ApJS..244....6S,2019AJ....158....5P,2020A&A...634A.102P}.

\subsection{Evolution and kinematics of components}
\begin{figure*}
\includegraphics[width=1\hsize]{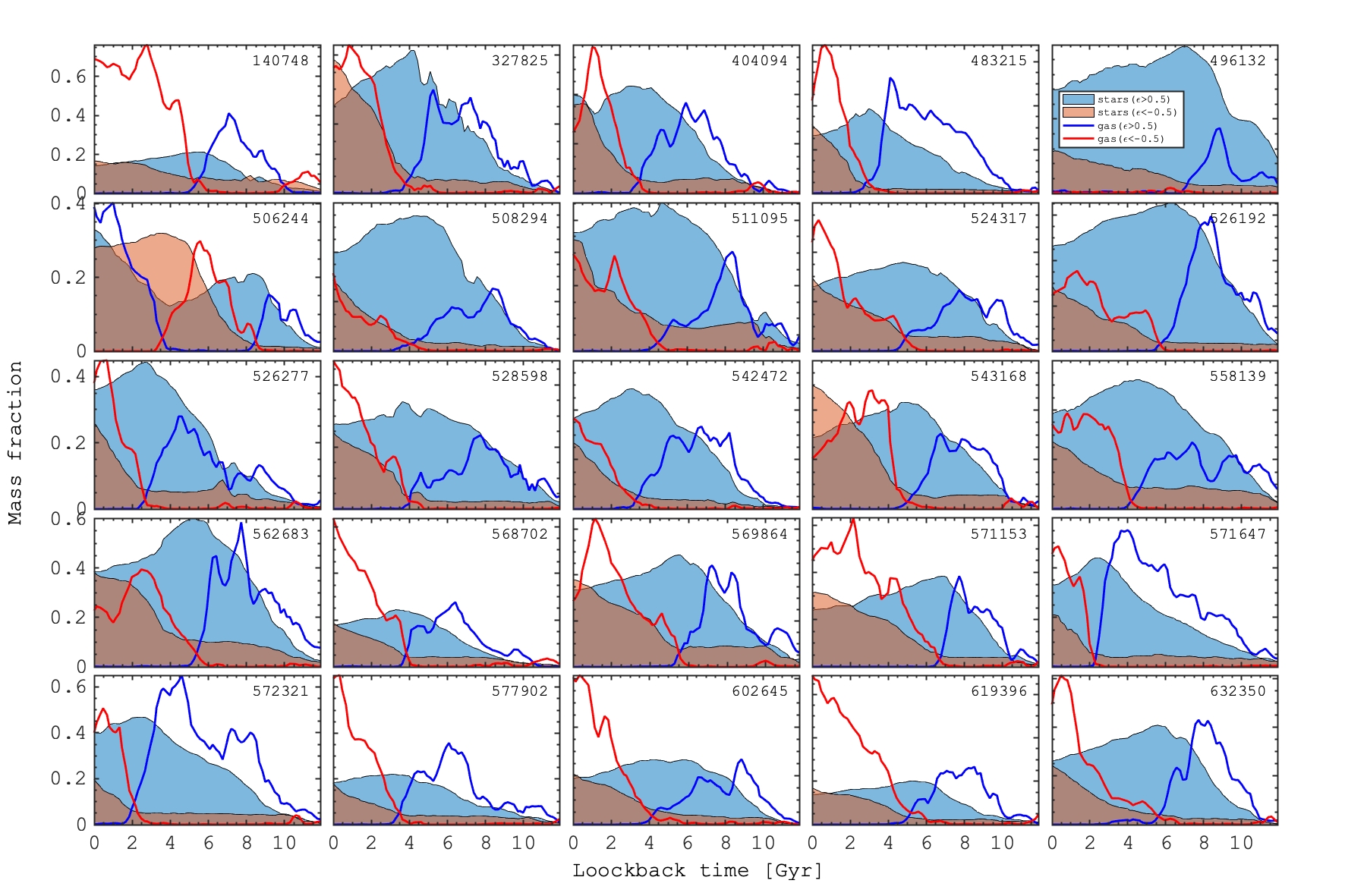}
\caption{Evolution of the mass fraction of gas~(lines) and stars~(filled) with $|\varepsilon|>0.5$ for the host~(blue) and counterrotating components~(red).}\label{fig::mass_evolution}
\end{figure*}

In this section we focus on the formation paths of galaxies with counterrotation in IllustrisTNG. In Fig.~\ref{fig::mass_evolution} we demonstrate the evolution of stellar (filled areas) and gas (solid lines) mass fraction. The stellar or gaseous components with the circularity $>0.5$~ are shown in blue and those with $<-0.5$~ in red. Such a circularity threshold allows us to follow the fractions that represent the disc components of the galaxies rather than stellar spheroidal components and gaseous halo. During the first stage~($6-8$~Gyr) the masses of gas and stars with positive circularity gradually increase~(blue), which suggests a smooth accretion of gas, likely from the circumgalactic medium~(CGM) feeding the star formation in the disc plane and, thus, leading to the formation of a dynamically cold host stellar component.

Later, the mass of the gas with $\epsilon>0.5$ slowly~(time scale depends on the galaxy and lies within the range: $1..3$~Gyr) decreases because of ongoing star formation~(with $\epsilon>0.5$). During this transition epoch, the galaxies acquire gas with negative circularity while the fraction of cold stars with positive circularity~($\epsilon>0.5$) decreases due to the enhancement of the velocity dispersion in the host component. As we already showed, the growth of the gas mass with negative circularity leads to the counterrotating stellar component formation. The star formation is efficient once the counterrotating gas settles down in the plane of the host stellar disc. An interesting feature here is that the amount of gas in the counterrotating component still increases due to a gradual accretion from the surrounding CGM.  Only in some galaxies in our sample it reaches a peak and then decreases due to its consumption by the star formation. The processes described above  are  observed in all our galaxies with counterrotation,  except the  140748, 506244 and may be 543168, where the direction of the gas rotation changed twice and lead to the formation of multiple misaligned components~(see also Fig.~\ref{fig::circularity_evolution} and Section~\ref{sec::global}).

\begin{figure*}
\includegraphics[width=1\hsize]{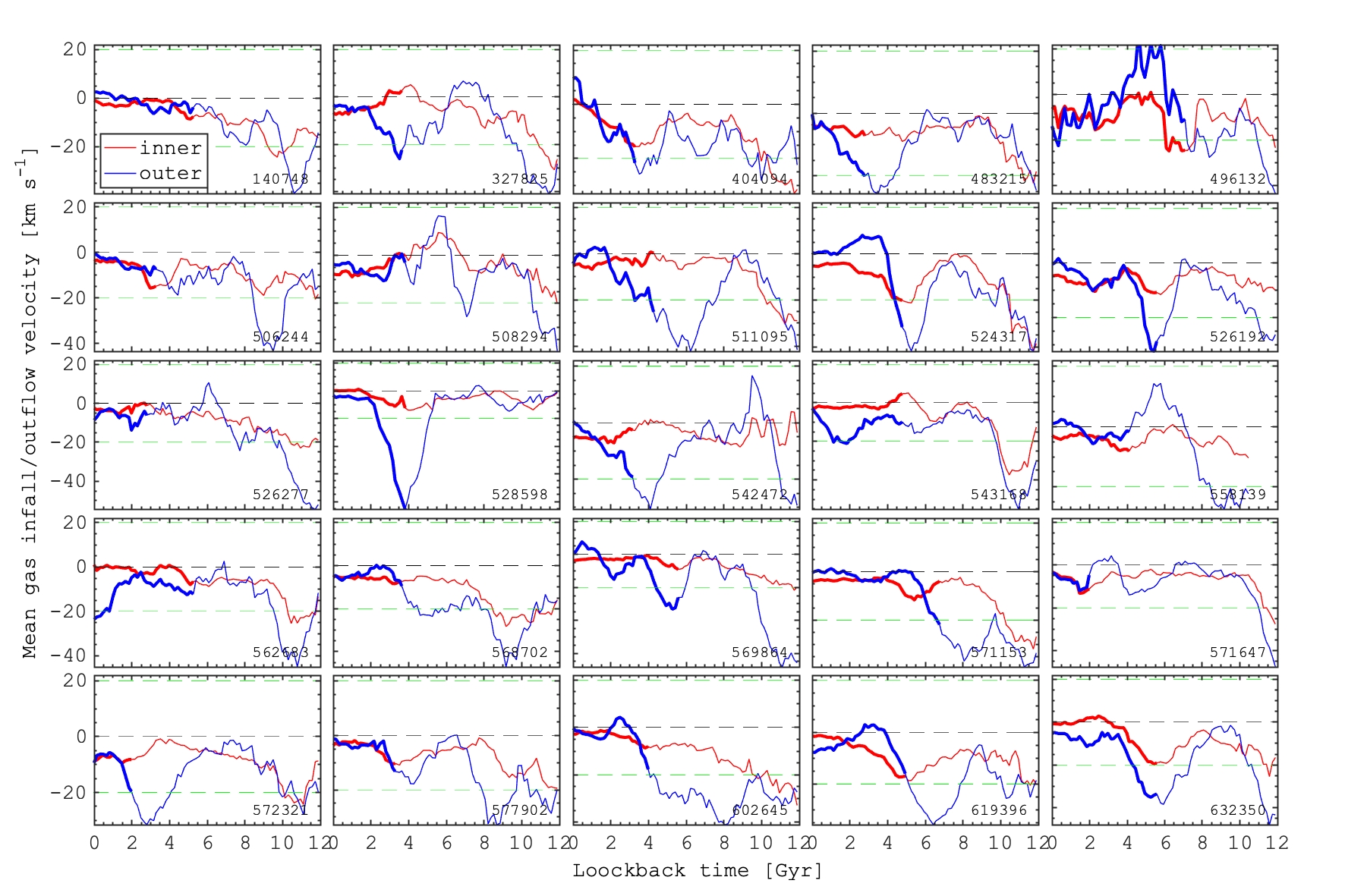}
\caption{Evolution of the mean gas velocity towards the galaxy~(radial velocity in spherical coordinates) in the galactic disc~(red, $|Z|<2$~kpc) and in the disc-halo interface~(blue, $2<|Z|<10$~kpc). 
Thick lines define the time after the gas circularity crosses $0$~(see Fig.~\ref{fig::circularity_evolution} for definition and Fig.~\ref{fig::accr_time} for the accretion time statistics).}\label{fig::gas_infall}
\end{figure*}

A slow decreasing of the gas amoint with positive circularity suggests that this gas is not removed from the discs by the feedback-driven outflows which are supposed to be efficient on a very short time scale. It seems that the gas in the host component is being replaced by the counterrotating one on a long time-scale of about $2-5$~Gyr. The infalling gas pushes out the existing gaseous component into the CGM, where both components mix together and then settle down to the disc plane on counterrotating orbits inherited from infalling gas. Mixing the pre-existing and infalling gas also explains why the metallicity of counterrotating stars is only slightly lower compared to the host stars at the outskirts, although at the end of evolution it is higher in the galactic center because of in-situ enrichment by new populations of counterrotating stars.

An important argument in favour of the pre-existing gas replacement is the absence of a gap between the end of the host component formation and the beginning of the counterrotating star formation. In particular, if the host gas { were removed by the feedback~(AGN or star formation burst) we would expect to observe a substantial gas-poor phase until a new~(retrograde)} gas approaches the galactic plane. On the contrary, the host gas removal and the counterrotating gas accretion are always synchronized thus revealing a manifestation of a single process resulting in the formation of the counterrotation.

\subsection{Gas accretion scenario}
To confirm our reasoning about the mechanism of the gas accretion/replacement, we investigate the kinematics of the gas at different distances from the galaxies. We expect that in case of the AGN-driven feedback, as it has been suggested by~\citep{2019ApJ...878..143S}, initially the gas velocity experiences perturbations near the disc plane and later, once the outflow propagates further away, in the outer regions. In the case of the external gas infall we expect to see the opposite: the outer parts of the gas would be perturbed first and then, once the accreted gas starts to interact with the disc, the velocity should be perturbed near the galactic plane. In Fig.~\ref{fig::gas_infall} we present the evolution of the mean mass-weighted radial gas velocity component~(in spherical coordinates) near the galactic plane $|z|<2$~kpc~(red) and at a larger distance $2<|z|<10$~kpc~(blue). Although the definition of the inner and outer galaxy is somewhat arbitrary, it still allows us to capture the main processes in the gas. Positive values of the velocity component correspond to the outflow while the negative velocity indicates an infall of gas. The bold parts of the curves correspond to the evolution after the flip~(from positive to negative values, see Fig.~\ref{fig::circularity_evolution} or Fig.~\ref{fig::mass_evolution}) of the mean gas circularity. The absolute values of the velocity do not correspond to the gas infall rate because they are obtained by averaging the velocity over the large area of the disc and surrounding medium thus provides only qualitative changes in the gas kinematics.

First, we do not observe any strong outflows~(positive radial velocity) from the inner galaxy~(red curves). Only 6 galaxies show a small positive velocity peaks near the moment of the gas reversal. Most of the galaxies however show an inflow~(or accretion) of gas from the outer regions which are associated with the negative velocity peaks~(blue). Indeed the amplitude, structure and temporal characteristics of the gas inflow in individual galaxies might be different especially taking into account some peculiarities in the orbital evolution of the infalling gas, which we have already demonstrated in Fig.~\ref{fig::PRGs}.  For instance, in the galaxy 506244, at early ages of~$\approx 9$~Gyr one can see an inflow of gas from the outer CGM, which corresponds to the first flip of the gas rotation shortly after~(see Fig.~\ref{fig::circularity_evolution} and Fig.~\ref{fig::mass_evolution}). Later, we observe an increase of the negative velocity at $4-5$~Gyr ago, which results in the formation of the second counterrotating component.

We conclude that the stellar counterrotation in our sample is the result of in-situ star formation~(see Fig.~\ref{fig::formation_radius_time_circularity}) from externally accreted, retrograde gas, which replaced a pre-existing gaseous disc. However, the relation between the misalignment~(and counterrotation) and the AGN activity~\citep[see, e.g.,][]{2019ApJ...878..143S,2020MNRAS.495.4542D} is still puzzling and we believe that it is important to uncover the physical processes standing behind such a picture. We address this issue in the next Section where we consider controlled simulations of retrograde and prograde gas infall onto galaxies, for better clarity.

\section{Controlled simulations of counterrotating galaxies formation with a gas replacement}\label{sec::SPH}
\subsection{Model description}
Unlike some of the previous merger/accretion simulations of counterrotating galaxies~\citep[see, e.g.,][]{1997ApJ...479..702T,2017MNRAS.471.1892B}, hereby we present our simulations of the gas accretion onto a gas-rich disc galaxy similar to ones we studied above from the TNG100 simulation. In our model the main galaxy is represented by the stellar-gaseous galactic disc. The gaseous subsystem is modelled with the Smoothed Particle Hydrodynamics, while the dynamics of the stellar disc is modelled by using a direct particle-particle integration scheme paralleled by the CUDA-GPU approach~\citep{Khrapov_Khoperskov_SPH_GPU_2017}. The number of particles is $2^{21}$ and $2^{22}$ for the stars and the gas, respectively, and the softening length is about $40$ pc for both components. The generation of the model initial equilibrium follows an iterative method of the Jeans equation solution~\citep{2011AstL...37..563K,2012MNRAS.427.1983K}. The simulation was performed with a direct GPU $N$-body/SPH code~\citep{Khrapov_Khoperskov_SPH_GPU_2017}, which we have tested in several galaxy dynamics studies where we obtained valid results with good performance and accuracy~\citep[see, e.g.,][]{Khrapov2019LagrangeEulerianMF,2020A&A...634L...8K}.

\begin{figure*}
\includegraphics[width=1\hsize]{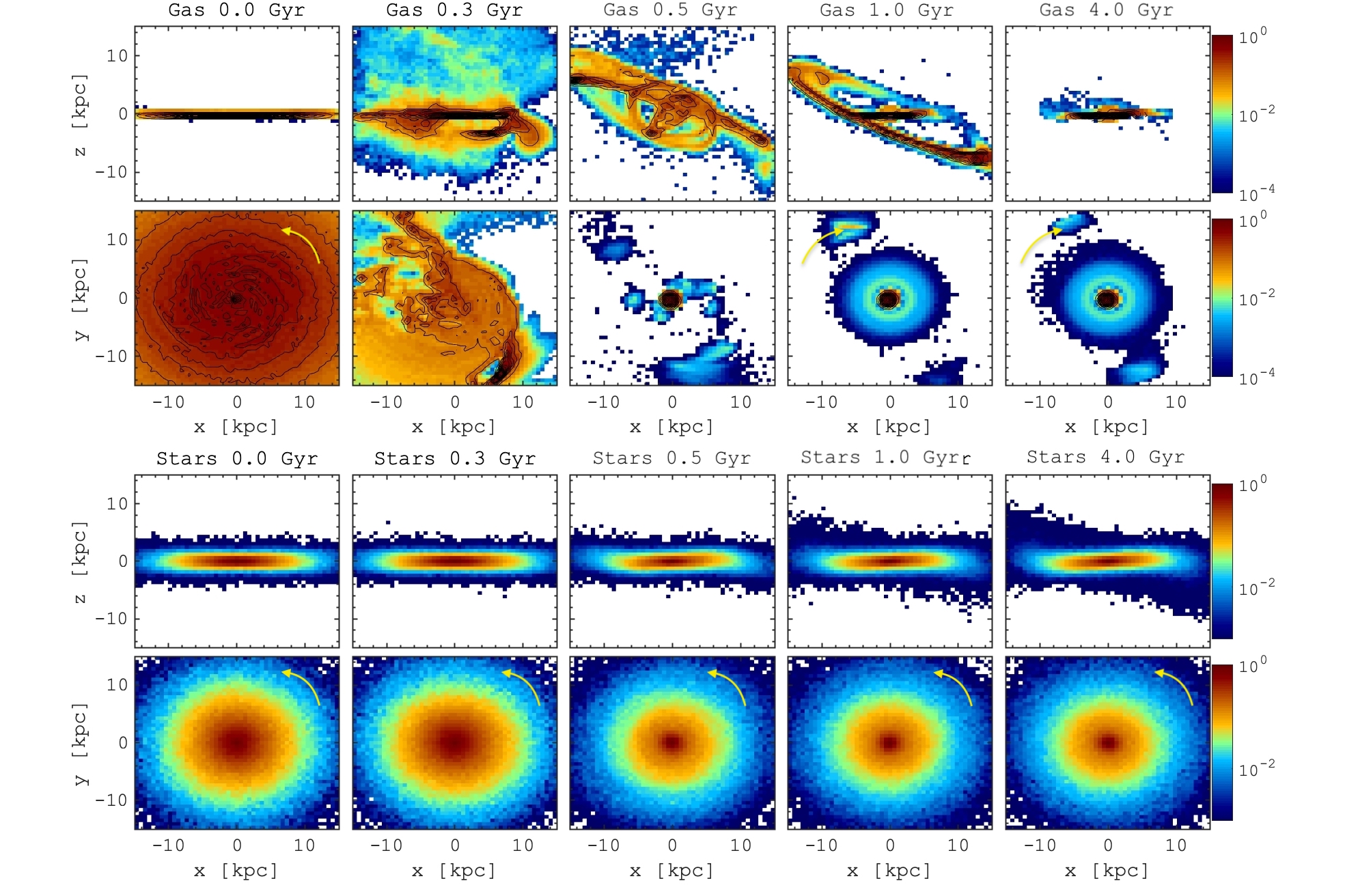}
\caption{Evolution of gas~(top frames) and stars~(bottom frames) in a simulation of a retrograde gas infall onto a gas-rich galaxy. For th ein-plane~(XY) maps we show only particles with $|z|<1$~kpc. The gas distribution maps include the gas with a temperature of $<20000$K. The galactic plane is always aligned with the XY-plane. The initial gaseous disc is kicked-out from the galactic plane~(stars) by the infalling gas, with almost a lack of gas in the disc plane at $\approx 0.5$. Once the gas settles back to the galactic plane, it inherits the direction of its rotation from the infalling gas. The stellar disc shrinks towards the center, but remains mostly unperturbed. The yellow arrows indicates the direction of rotation of gas and stars.}\label{fig::sph_evol}
\end{figure*}


Initially, the infalling material is represented by six large gaseous clumps moving along a parabolic trajectory inclined by a $30^\circ$ with respect to the galactic plane with the initial velocity of $0.8$ of the circular value. All extragalactic gas is located beyond the optical size of the galaxy~($25$~kpc) at that time. The mass of infalling gas is twice larger than the gas in the main galaxy, which is only $10\%$ of the total galaxy mass. Once the external gas starts moving towards the galaxy, it rapidly stretches along the trajectory, and forms a stream or filament-like structure soon. In this paper we focus on two models, where the direction of the gas infall is retrograde and prograde with respect to the initial rotation of the main galaxy. Although the parameters of the models may not represent the systems we analyze above in their entirety, we are confident that the general patterns of evolution and formation of the counterrotating system are well described by our simplified modelling. Note that our modelling includes neither the AGN-driven feedback nor the star formation. This allows us to focus on the dynamical interaction between the infalling gas and the gas-rich disc galaxy. In this paper we present only two models, while the detailed investigation of the relative mass impact, the infall orbital parameters study and impact of feedback is a subject of a follow-up work~(Khrapov et al. in prep).

\subsection{The gaseous disc replacement via a retrograde accretion}
\begin{figure}
\includegraphics[width=1\hsize]{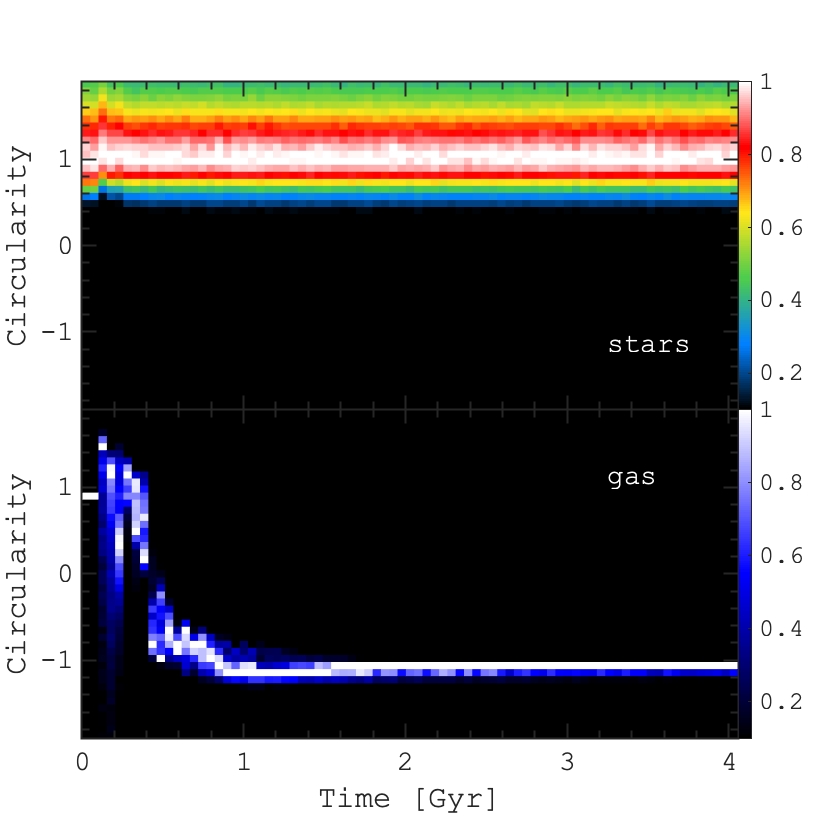}
\caption{Evolution of the circularity distribution for stars~(top) and gas~(bottom) in a toy model of the retrograde gas infall onto a gas-rich galaxy, similar to Fig.~\ref{fig::circularity_evolution} for the IllustrisTNG galaxies).}\label{fig::sph_circularity}
\end{figure}

In Fig.~\ref{fig::sph_evol} we show the evolution of both stellar and gaseous components in our retrograde accretion model. At the end of our simulation only a small fraction of the infalling gas can be still found along its initial orbit, while the most of it is perfectly aligned with the stellar disc. The stellar disc, in fact, does not experience any significant perturbation, but it shrinks instead. This is consistent with the finding that the host stellar discs are usually shorter compare to counterrotating components among the Illustris galaxies~(see Fig.~\ref{fig::circularity_radius} and \ref{fig::mass_fraction}). The dynamics of the system can be traced in more detail in Fig.~\ref{fig::sph_circularity}, where we show the evolution of circularity, similarly to Fig.~\ref{fig::circularity_evolution}, where we also distinguish the gas initially present in the galaxy from the infalling material. We can see that the evolution of our simulated galaxy looks rather similar to the Illustris ones. The accreted gas obviously settles down in the galactic plane and forms a new retrograde gaseous disc. The only mechanism able to change the spin of the gas is that it has been kicked out from the disc and trapped by the infalling gas. Therefore, the resulting gaseous disc actually is a mixture of accreted gas and pre-existing gas reversed during the infall. Also we note that the stellar disc experiences some modest heating which we have also found in the Illustris galaxies. Therefore, we can conclude that the scenario we propose for the formation of galaxies with counterrotation is preferable. In particular, a retrograde gas accretion is able to replace the gaseous component, and once it settles down in the stellar disc plane, it is able to form stars contributing to the counterrotating stellar component. 

\subsection{Feeding the galaxies: retrograde versus prograde gas infall}
\begin{figure}
\includegraphics[width=1\hsize]{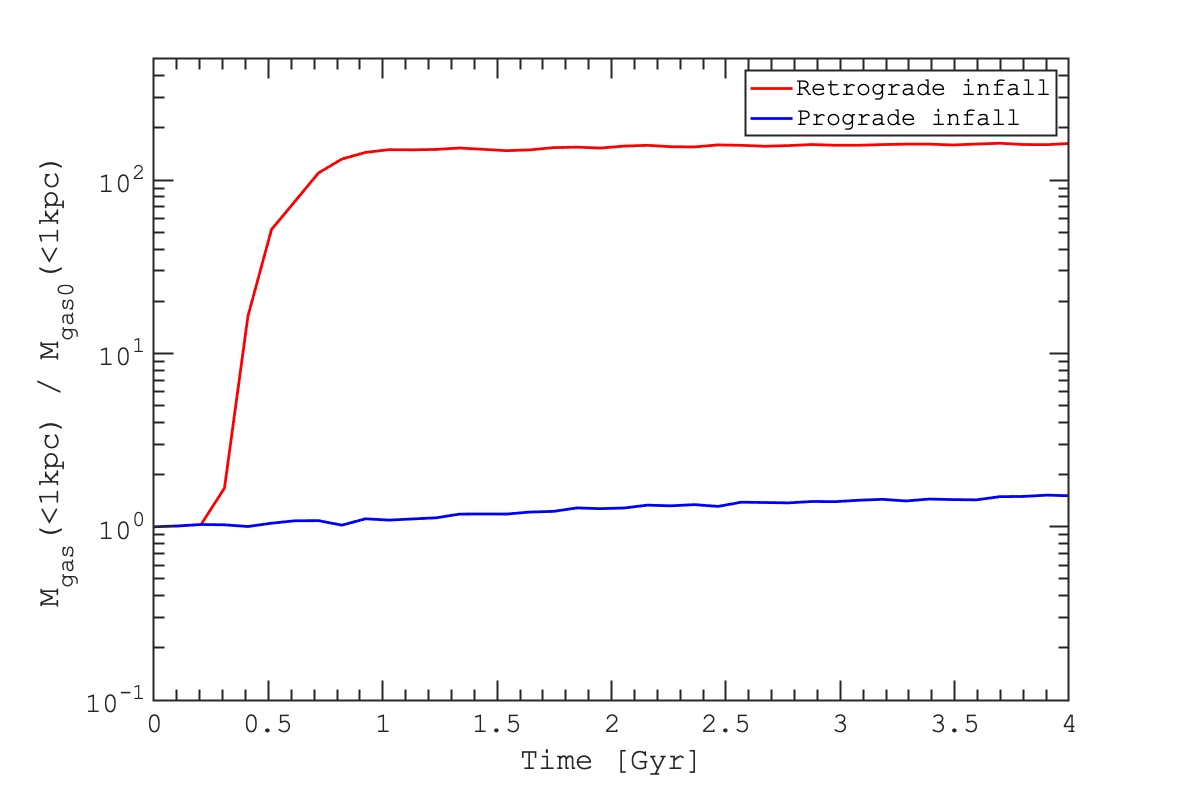}
\caption{The gas mass in the central kpc in two toy models of the retrograde~(red) and prograde~(blue) gas infall to the gas-rich galaxy. During the retrograde accretion the gas efficiently falls into the galactic center and is able to stimulate the activity of a central black hole.}\label{sph::accretion}
\end{figure}

Below we discuss a possible link between the misaligned~(counterrotation is the extreme case) components in galaxies and a central black hole activity. Since our models do not include the black hole evolution, we focus on the evolution of gas in the central region, potentially available for the central black hole feeding. In Fig.~\ref{sph::accretion} we show the evolution of the gas mass in the central $1$ kpc in the model described above, in comparison to a similar model, but with a prograde accretion. We find an interesting behaviour: during its very first encounter the retrograde gas, while being unable to pass through, sticks together with the gas existing in the galaxy, loses its angular momentum and rapidly  falls to the galactic center. At the same time, in the case of a prograde gas infall, the galactic center feeding is less efficient by a factor of $\sim 10^2$ which suggests also a less efficient black hole activity. Since the retrograde gas infall onto the galaxy causes a more efficient feeding of the galactic centre with respect to a prograde infall, it is obvious that the presence of misaligned components correlates with an AGN activity, as it has been noticed in the previous works~\citep{2019ApJ...878..143S,2020MNRAS.495.4542D}. Therefore, we suggest that the AGN activity is efficiently enhanced by the formation of misaligned components via retrograde gas infall, but it does not cause the misalignment.

\section{Conclusions}\label{sec::conclusion}
In this paper we present an in-depth analysis of galaxies with counterrotation from the TNG100 simulation. Recent progress of the cosmological simulations allowed us to focus on the internal kinematics, ages, metallicities, and structure, as well as on the origin of stellar and gaseous components of galaxies with counterrotation. Our conclusions are based on the exploration of 25 galaxies selected at $z=0$ and can be summarized as follows:

\begin{itemize}
\item[1.] By analysing the circularity distributions, we find $25$ galaxies with pairs of embedded stellar discs rotating in the same plane but in different directions, thus providing classical examples of star-star counterrotating systems. Although our sample selection is based on the stellar counterrotation, we find that all the galaxies with counterrotating stellar discs also harbour cold gaseous discs~(see Figs.~\ref{fig::circularity0},~\ref{fig::circularity_radius}).

\item[2.] From our analysis of the gas kinematics in the galaxies and in nearby CGM together with circularity evolution, we conclude that stellar counterrotation in simulated galaxies is the result of in-situ star formation~(see Fig.~\ref{fig::formation_radius_time_circularity}) in the counterrotating gaseous component accreted from an external reservoir~(see Fig.~\ref{fig::gas_infall}). This counterrotating gaseous component replaces the pre-existing co-rotating gaseous disc during the external gas accretion~(see Fig.~\ref{fig::circularity_evolution}). Therefore, an important stage in the formation of galaxies with counterrotation is a significant gas loss followed by the accretion of material with misaligned angular momentum. Since we do not find any temporal gaps between the removal of co-rotating gas and acquisition of counterrotating gas, we suggest that the co-rotating pre-existed gaseous disc has been pushed out from the galaxy by the infalling component. A major merger scenario can also be ruled out because we do not find a presence of an accreted stellar component, except for a possible very gas-rich dwarf accretion. Therefore, the most favourable source of the misalignment gas is the cosmological  filaments or gas-rich CGM in the outskirts of nearby massive galaxies.
 
\item[3.] The time scale of the gas accretion has a log-normal distribution that peaks at $\approx 4$~Gyr ago. During the gas infall, the misaligned components can be found as extended discs or rings randomly orientated relative to the host stellar component~(see Fig.~\ref{fig::infall_angle}). Surprisingly, some of the counterrotating galaxies look like polar ring galaxies at the time of the gas accretion, which suggests, for the first time, a possible genetic connection between these two types of multi-spin galaxies~(see Fig.~\ref{fig::PRGs}).

\item[4.] The fraction of counterrotating stars in our sample is almost constant across the galaxies with an enhancement towards the larger galactocentric distances, which demonstrates that counterrotating components are more radially extended than the hosts.  Counterrotating and host components are clearly separated by the mean stellar age. In particular, we find that the counterrotating stars are on average $\approx2-4$~Gyr younger than the host stars. The metallicity distributions in both stellar components are rather similar, however in the outer parts the counterrotating components are slightly more metal-poor, while in the centre we observe the opposite behaviour~(see Figs.~\ref{fig::mass_size_compare}, ~\ref{fig::age_met}). Such a picture can be explained by the accretion of external gas through the tidal interaction with a gas rich donor galaxy or from a cosmological filament. The low-metallicity in the outer counterrotating component is explained by the dilution of metals, while the enrichment in the galactic centre is a result of increased star formation activity.

\item[5.] The counterrotating galaxies demonstrate a distinct feature in the velocity dispersion distribution. In agreement with some previous observational studies, we find that both stellar components of counterrotating galaxies tend to be vertically hotter compared to normal galaxies. The vertical velocity dispersion is higher by $20-40$~\kmps, leading to atypical values of the vertical-to-radial velocity dispersion ratio $\geq 1$ across the entire galaxy~(see Figs.~\ref{fig::mean_dispersion},~\ref{app}). Such a behaviour is likely to be a result of the bending instability predicted by theoretical studies~\citep{1994ApJ...425..551M,2017A&A...597A.103K}. This result suggests that the presence of misalignment components in galaxies can be an important source of the discs heating over time. Since cosmological models of the galactic assembly predict the accretion of multiple objects at different epoch, the presence of misaligned components should be ubiquitous in galaxies, which in turn should contribute to the velocity dispersion anisotropy in normal galaxies.

\item[6.] We find that the retrograde gas infalling onto the gas-rich galaxy is responsible for the formation of large-scale counterrotation on a longer time scale, and leads to an efficient feeding of the central region of the galaxy, which may be responsible for the subsequent AGN activity. Therefore, we conclude that the AGN activity can be a signature of a misaligned formation via an external gas accretion, but the reversal is not true: it does not cause the appearance of the counterrotating components.

\end{itemize} 

\section{Data availability}
Data directly related to this publication and its figures are available by request from the corresponding author. The IllustrisTNG simulations themselves are publicly available and accessible at \url{www.tng-project.org/data}~\citep{2019ComAC...6....2N}.

\section*{Acknowledgements}
We thank the anonymous referee for the rapid report and for helpful comments that improved the clarity of the paper. We also thank Tjitske Starkenburg for useful discussions. The analysis of simulations and new models have been performed by using the equipment of the shared research facilities of HPC computing resources at Lomonosov Moscow State University (project RFMEFI62117X001). AS and SK acknowledge the Russian Science Foundation (RSCF) grant 19-72-20089. IAZ acknowledges support by the grant for young scientist's research laboratories of the National Academy of Sciences of Ukraine.  SSK and AVK thank the Ministry of Science and Higher Education of the Russian Federation (government task no. 0633-2020-0003, the results of numerical simulations of galaxies dynamics in section 6). MI acknowledges support by the National Academy of Sciences of Ukraine under the Research Project of young scientists No. 0119U102399. PB acknowledges support by the Chinese Academy of Sciences  through the Silk Road Project at NAOC, the President's International Fellowship (PIFI) for Visiting Scientists program of CAS, the National Science Foundation of China under grant No. 11673032. This work was supported by the Deutsche Forschungsgemeinschaft (DFG, German Research Foundation) – Project-ID 138713538 -- SFB 881 ('The Milky Way System'), by the Volkswagen Foundation under the Trilateral Partnerships grants No. 90411 and 97778. The work of PB and MI was supported under the special program of the NRF of Ukraine ``Leading and Young Scientists Research Support'' - "Astrophysical Relativistic Galactic Objects (ARGO): life cycle of active nucleus",  No. 2020.02/0346. The reported study was partially funded by RFBR and DFG according to the research project 20-52-12009. BA acknowledges support by the \emph{“Landesgraduiertenstipendium"} of the University of Heidelberg and the Volkswagen Foundation under the Trilateral Collaboration Scheme (Russia, Ukraine, Germany) project titled ("Accretion Processes in Galactic Nuclei").

{\it Software:} \texttt{IPython}~\citep{2007CSE.....9c..21P}, \texttt{Astropy} \citep{2013A&A...558A..33A, 2018AJ....156..123A}, \texttt{NumPy} \citep{2011CSE....13b..22V}, \texttt{SciPy} \citep{2020SciPy-NMeth}, \texttt{AGAMA} \citep{2019MNRAS.482.1525V}, \texttt{Matplotlib} \citep{2007CSE.....9...90H}, \texttt{Pandas} \citep{mckinney-proc-scipy-2010}, \texttt{Illustris-TNG JupyterLab interface}~\citep{2019ComAC...6....2N}.

\bibliographystyle{mnras}
\bibliography{paper_illustris.bib}

\appendix

\section{Extra plots}\label{app}

\begin{figure}
\includegraphics[width=1\hsize]{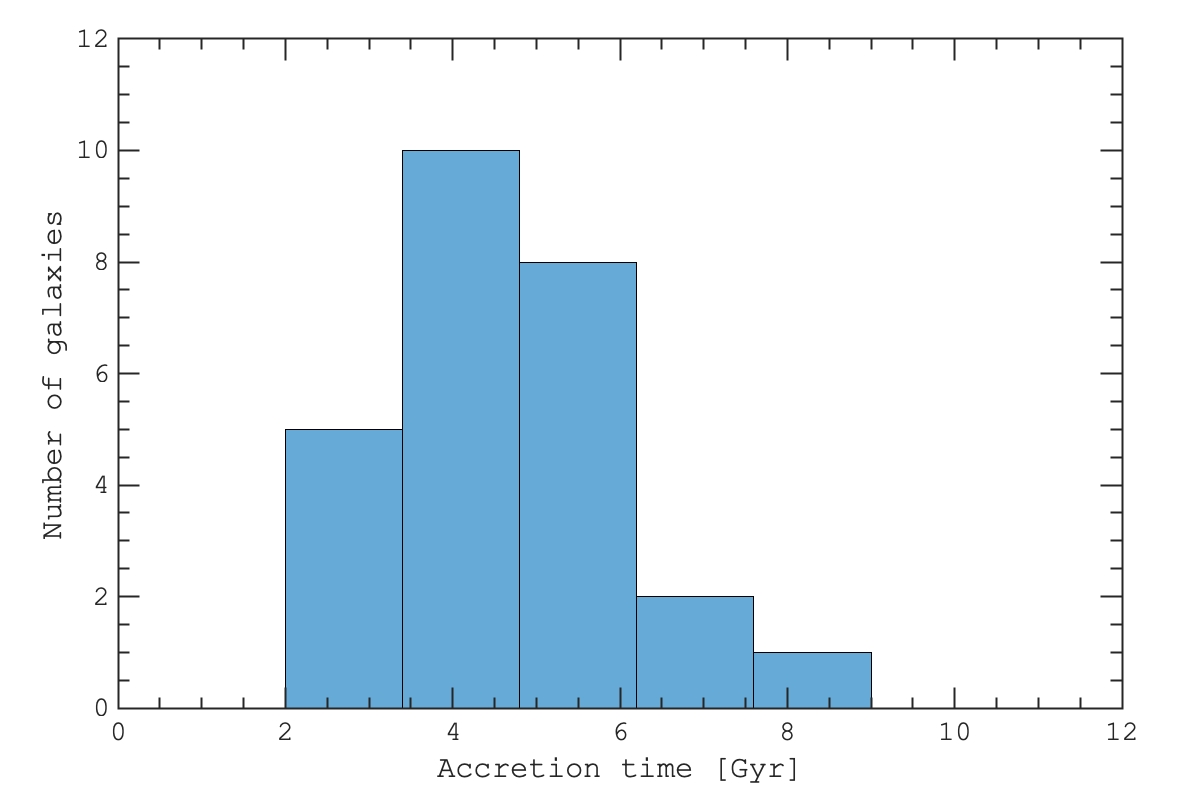}
\caption{Distribution of the accretion time~($\rm T_{accr}$) in galaxies with counterrotation identified in Fig.~\ref{fig::mass_evolution} as a transition between a positive and negative circularity of the gaseous component.}\label{fig::accr_time}
\end{figure}


\begin{figure*}
\includegraphics[width=0.5\hsize]{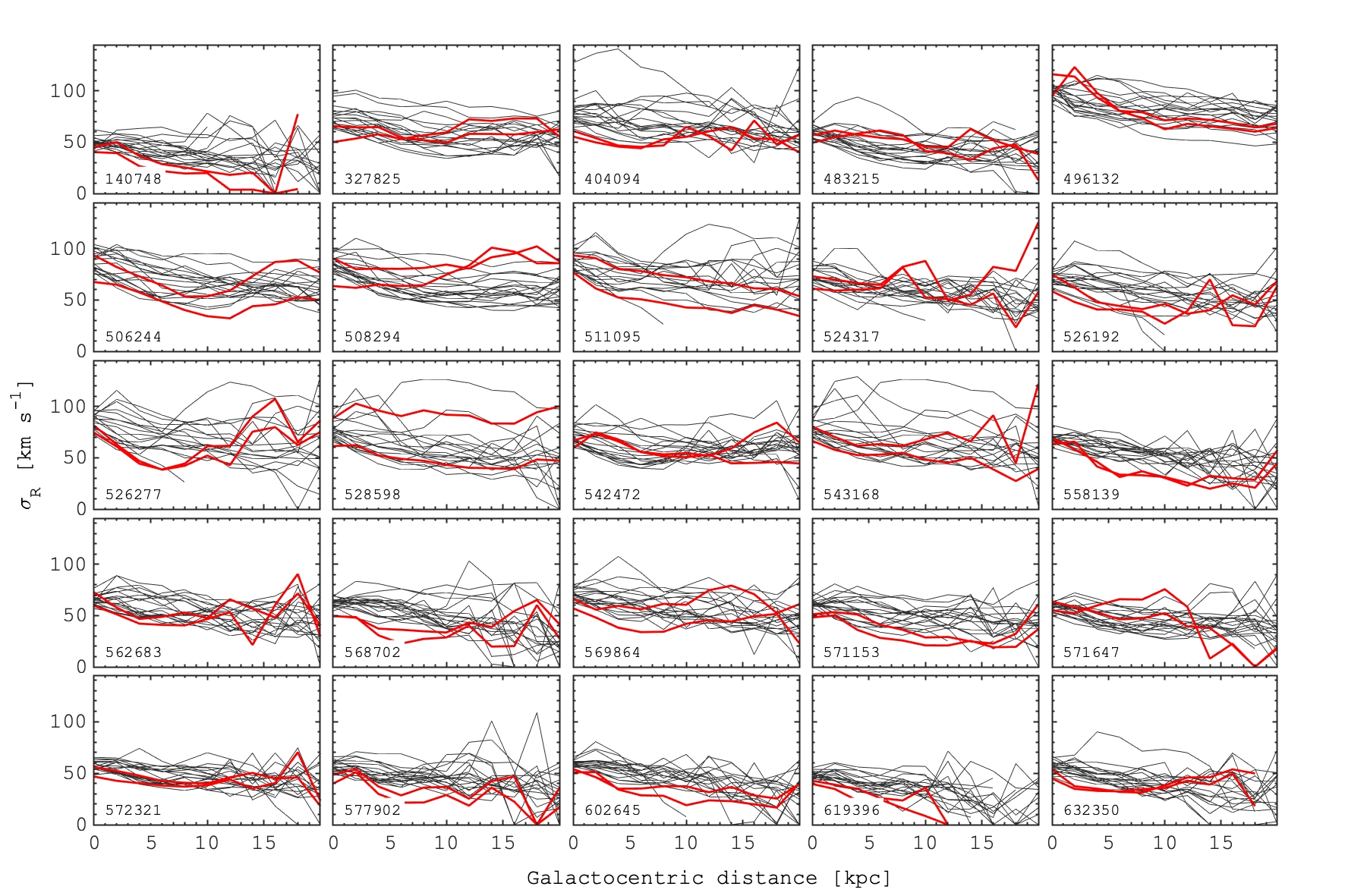}\includegraphics[width=0.5\hsize]{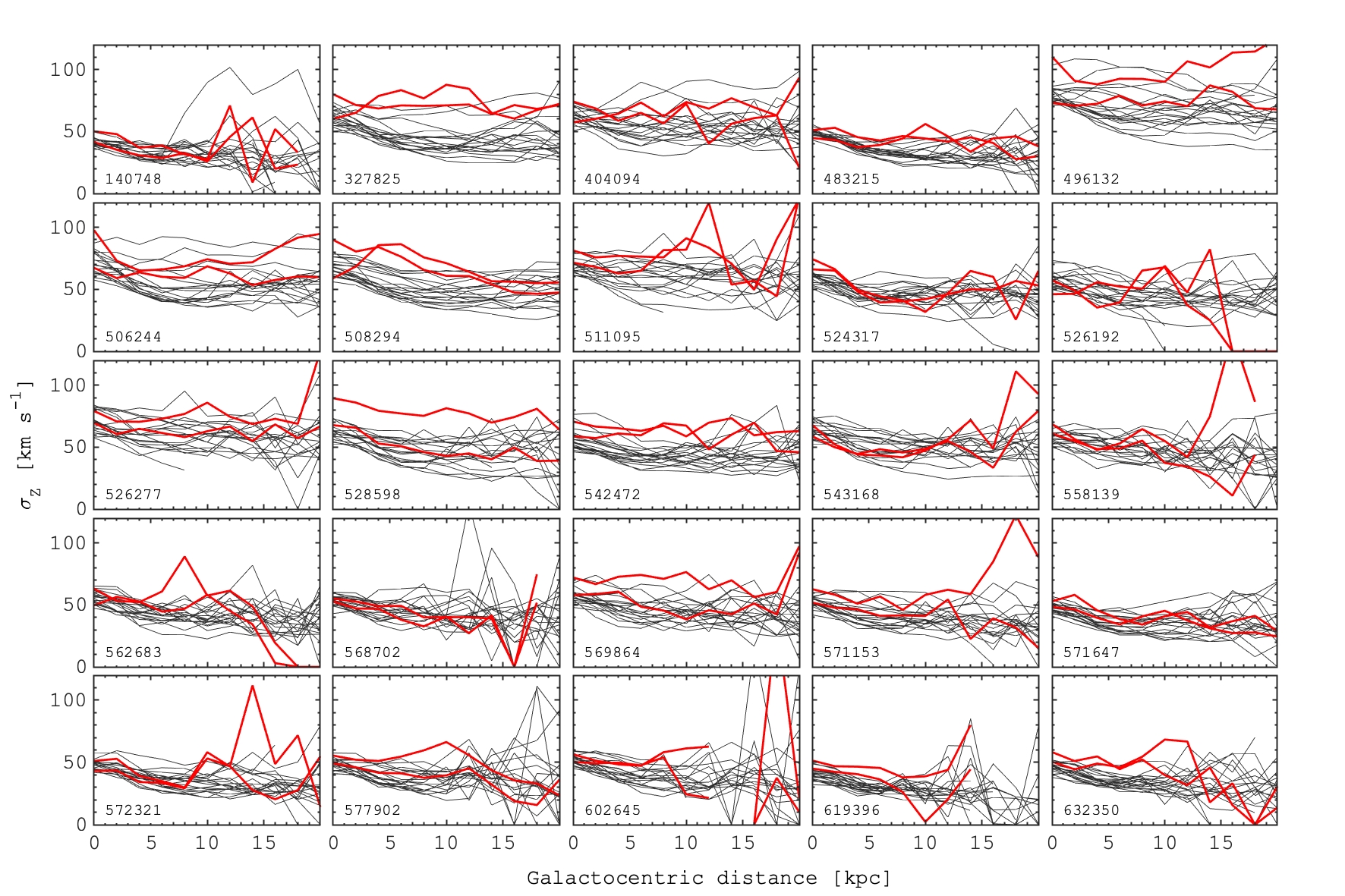}
\caption{{\it Left:} Radial velocity dispersion profiles in galaxies with counterrotation~(each component is shown separately by the red lines) together with the comparison sample of 20 different galaxies chosen according to the mass and size of the galaxy with counterrotation. {\it Right:} same as in left but for the vertical velocity dispersion profiles.}\label{fig::app_cr}
\end{figure*}

\begin{figure*}
\includegraphics[width=0.5\hsize]{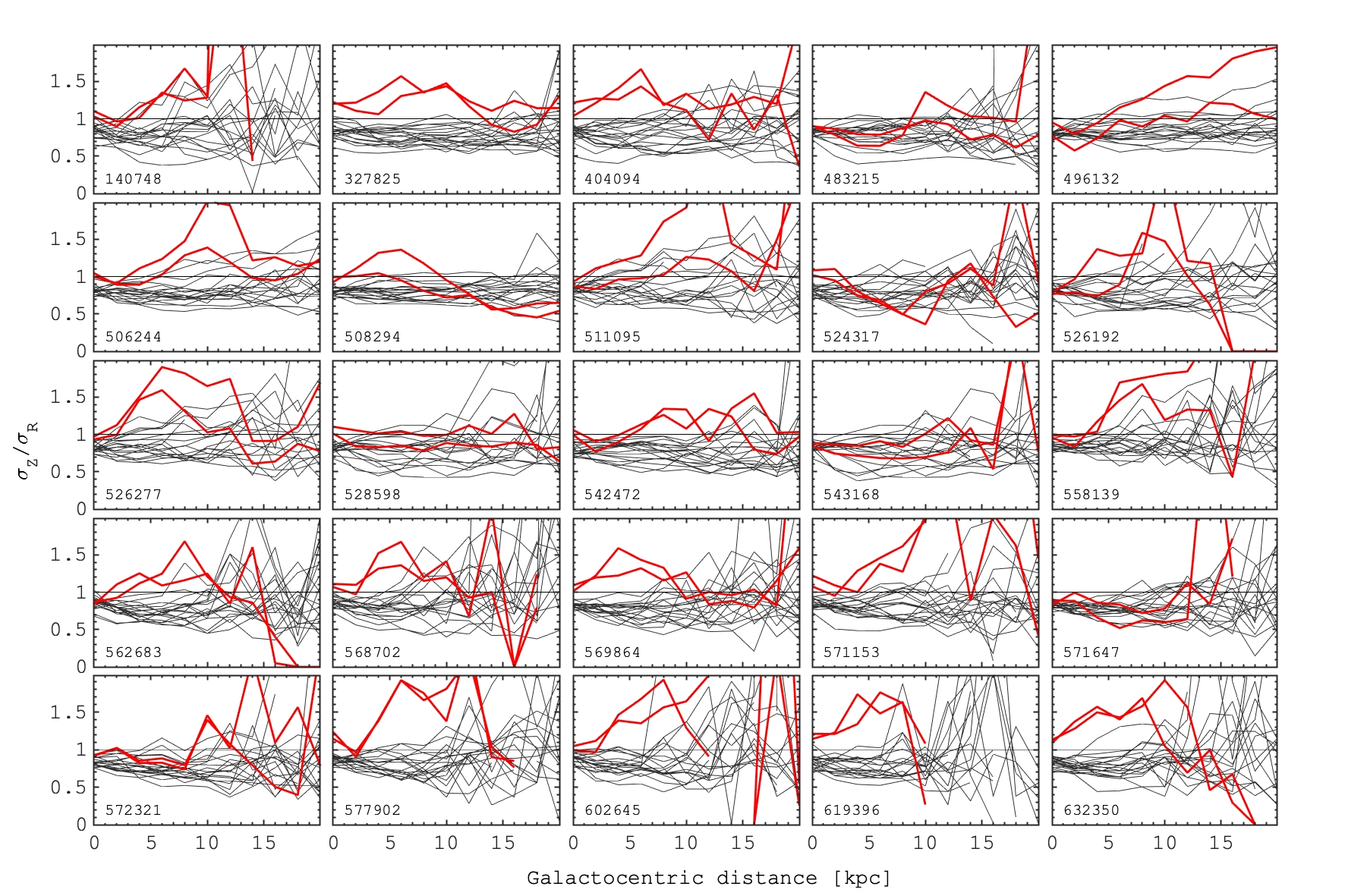}
\caption{Same as in Fig.~\ref{fig::app_cr} but for the vertical to radial velocity dispersion ratio. }\label{fig::app_czcr}
\end{figure*}

\bsp	
\label{lastpage}
\end{document}